\documentclass[a4paper,fleqn,usenatbib]{mnras}

\usepackage[T1]{fontenc}
\usepackage{ae,aecompl}

\usepackage[pdftex]{graphicx}
\usepackage{amsmath}    
\usepackage{amssymb}    

\title[Discovery of two embedded massive YSOs and an outflow in IRAS~18144-1723]{Discovery of two embedded massive YSOs and an outflow in IRAS~18144-1723}

\author[W. P. Varricatt et al.]{
W. P. Varricatt$^{1}$\thanks{E-mail: w.varricatt@ukirt.hawaii.edu},
J. G. A. Wouterloot$^{2}$,
S. K. Ramsay$^{3}$,
and C. J. Davis$^{4,5}$
\\
$^{1}$Institute for Astronomy, UKIRT Observatory, 660 N. A\'{o}hoku place, Hilo, HI - 96720, USA\\
$^{2}$East Asian Observatory, 660 N. A\'{o}hoku Place, University Park, Hilo, HI - 96720, USA\\
$^{3}$European Southern Observatory, Karl-Schwarzschild-Str. 2, 85748 Garching bei M\"{u}nchen, Germany\\
$^{4}$Astrophysics Research Institute, Liverpool John Moores University, 146 Brownlow Hill, Liverpool L3 5RF\\
$^{5}$National Science Foundation, Division of Astronomical Sciences, 2415 Eisenhower Avenue, Alexandria, VA 22314 (Current address)\\
}

\date{Accepted 2018 July 30; Received 2018 July 27; in original form 2017 September 13}

\pubyear{2018}

\begin{document}
\label{firstpage}
\pagerange{\pageref{firstpage}--\pageref{lastpage}}
\maketitle

\begin{abstract}
Massive stars are rarely seen to form in isolation. It has been
proposed that association with companions or clusters in the 
formative stages is vital to their mass accumulation. In this 
paper we study IRAS~18144-1723, a massive young stellar
object (YSO) which had been perceived in early studies as a 
single source. 
In the CO(3-2) line, we detect an outflow aligned 
well with the outflow seen in H$_2$ in this region. We show 
that there are at least two YSOs here, and that the outflow is 
most likely to be from a deeply embedded source detected in 
our infrared imaging. Using multi-wavelength observations, 
we study the outflow and the embedded source and derive their 
properties. 
We conclude that IRAS~18144 hosts an isolated cloud, in 
which at least two massive YSOs are being born.
From our sub-mm observations, we derive
the mass of the cloud and the core hosting the YSOs.
\end{abstract}




\section{Introduction}

Recent studies suggest that the primary mechanism for the formation
of massive stars is disk accretion as in their low-mass counterparts
\citep{Arce07, Beuther02a, Varricatt10, Davis04}. Collimated 
outflows discovered from massive star forming regions in these studies 
indicate the presence of accretion discs.
The cavities carved by the protostellars outflow provide a path
for the radiation to escape, thereby reducing
the radiation pressure on the accreted matter and allowing
the mass accumulation in massive YSOs through accretion \citep{Krumholz05}.
However, unlike low-mass stars,
massive stars are rarely observed to form in isolation. Instead, they 
are seen to be associated with companions or clusters suggesting that 
such associations may play a prominent role in their formation \citep{Lada03}.
Massive YSOs are
located in the galactic plane at large distances from us, so we need
observations at infrared and longer wavelengths at high angular
resolution to understand their formation.  With the availability of
large telescopes operating at these wavelengths, many of the massive 
YSOs, which appeared to form in isolation in 
older studies, are now being resolved into binaries or multiples
(e.g. \citet{Varricatt10}).
In this paper, we present a detailed observational study of the massive 
YSO IRAS~18144-1723 (hereafter IRAS~18144).

IRAS~18144 is located near the galactic plane 
(l=13.657$^{\circ}$, b=-0.6$^{\circ}$), and is associated with 
a dense core of far-IR luminosity 1.32$\times$10$^4$~L$_\odot$, 
detected in NH$_3$ emission \citep{Molinari96}. Using the 
line-of-sight radial velocity of the NH$_3$ emission 
(+47.3\,km~s$^{-1}$), they derived a kinematic distance 
of 4.33\,kpc. Water and methanol masers are also detected from 
this source \citep{Palla91, Szymczak00, Kurtz04, Gomez-Ruiz16}.
\citet{Molinari98} did not detect 6-cm radio emission 
associated with the IRAS source sugessting that it is
in a pre-UCH{\sc{ii}} stage.
Observations by \citet{Zhang05}
did not detect any CO(2--1) outflow towards this region. However, through near-IR
imaging in the H$_2$ 
line at 2.1218\,$\mu$m and in the $K$ filter, \citet{Varricatt10}
discovered an E--W jet appearing to emanate from a bright near-infrared
source (referred to as `A') located at $\alpha$=18:17:24.38, $\delta$=-17:22:14.7\footnote{All 
coordinates given in this paper are in J2000}.
We further observed this region at multiple wavelengths from near-IR
to sub-mm in order to
uncover the nature and possible multiplicity of source `A'.
The new observations are analysed along with archival data.

\section[]{Observations and data reduction}

\subsection[]{Near-IR imaging with WFCAM}

We observed IRAS~18144 using the United Kingdom Infrared 
Telescope (UKIRT), Hawaii, and the UKIRT Wide Field Camera 
(WFCAM; \cite{Casali07}). WFCAM employs four 2048$\times$2048 
HgCdTe HawaiiII arrays, each with a field of view of 
13.5\arcmin$\times$13.5\arcmin at an image scale of 
0.4\arcsec~pixel$^{-1}$.

Observations were obtained in the the near-IR $J, H$ and 
$K$-band filters, and in narrowband filters centred on the 
H$_2$ $\upsilon$ = 1--0 S(1) line at 2.1218~$\mu$m and the 
[Fe {\sc{II}}] $a^4 D_{7/2} - a^4 F_{9/2}$  forbidden emission 
line at 1.6439~$\mu$m.
H$_2$ and [\ion{Fe}{ii}] are good tracers of star formation
activity. H$_2$ emission may be fluorescently excited in 
photodissociation regions associated with massive and 
intermediate mass young stars or in planetary nebulae,
or shock excited in jets and warm entrained 
molecular outflows from YSOs. [Fe {\sc{II}}] 
is often used as a tracer of
ionized region at the base of the jet \citep{Davis11}. 

\begin{table}
\centering
\caption{Imaging observations using WFCAM and UIST}
\label{tab:UKIRT_imaging_log}
\begin{tabular}{@{}lllll@{}}
\hline
UTDate    & Filters & Exp. time &Int. time &FWHM      \\
yyyymmdd  &         & (sec)     &(sec)     &(arcsec)       \\
\hline
\multicolumn{5}{c}{WFCAM}\\
\hline
20140620  & $J$     & 10,2   &360, 40  & 1.11, 1.10     \\
20140620  & $H$     & 5,1    &180, 20  & 1.10, 1.21     \\
20140620  & $K$     & 5,1    &180, 20  & 1.05, 0.91     \\
20140620  & 1-0S1   & 40     &1440  & 1.07       \\
20140621  & 1-0S1   & 40     &2$\times$1440  & 0.82, 0.79       \\
20140621  & $J$ & 10        &360            & 0.84       \\
20140621  & $H,K$ & 5,5   &180, 180   & 0.78, 0.77       \\
20140622  & 1.644FeII & 40   &1440  & 0.79       \\
20140623  & 1.644FeII & 40   &1440  & 0.86       \\
20170522  & $J$     & 5      &720    & 0.99   \\
20170522  & $H, K$  & 5, 5   &360, 180    & 0.82, 0.84   \\
20180330  & $J$     & 5      &720    & 0.94   \\
20180330  & $H, K$  & 5, 5   &360, 180    & 0.88, 0.87   \\
\hline
\multicolumn{5}{c}{UIST}\\
\hline
20140820  & $L'$ &0.4$\times$50$^{\mathrm{a}}$     &160               &0.48      \\
20140820  & $M'$ &0.175$\times$50    &105               &0.5       \\
\hline
\end{tabular}
\begin{list}{}{}
\item {$^{\mathrm{a}}$exp. time $\times$ coadds}
\end{list}
\end{table}

\begin{figure*}
\centering
\includegraphics[width=17.7cm,clip]{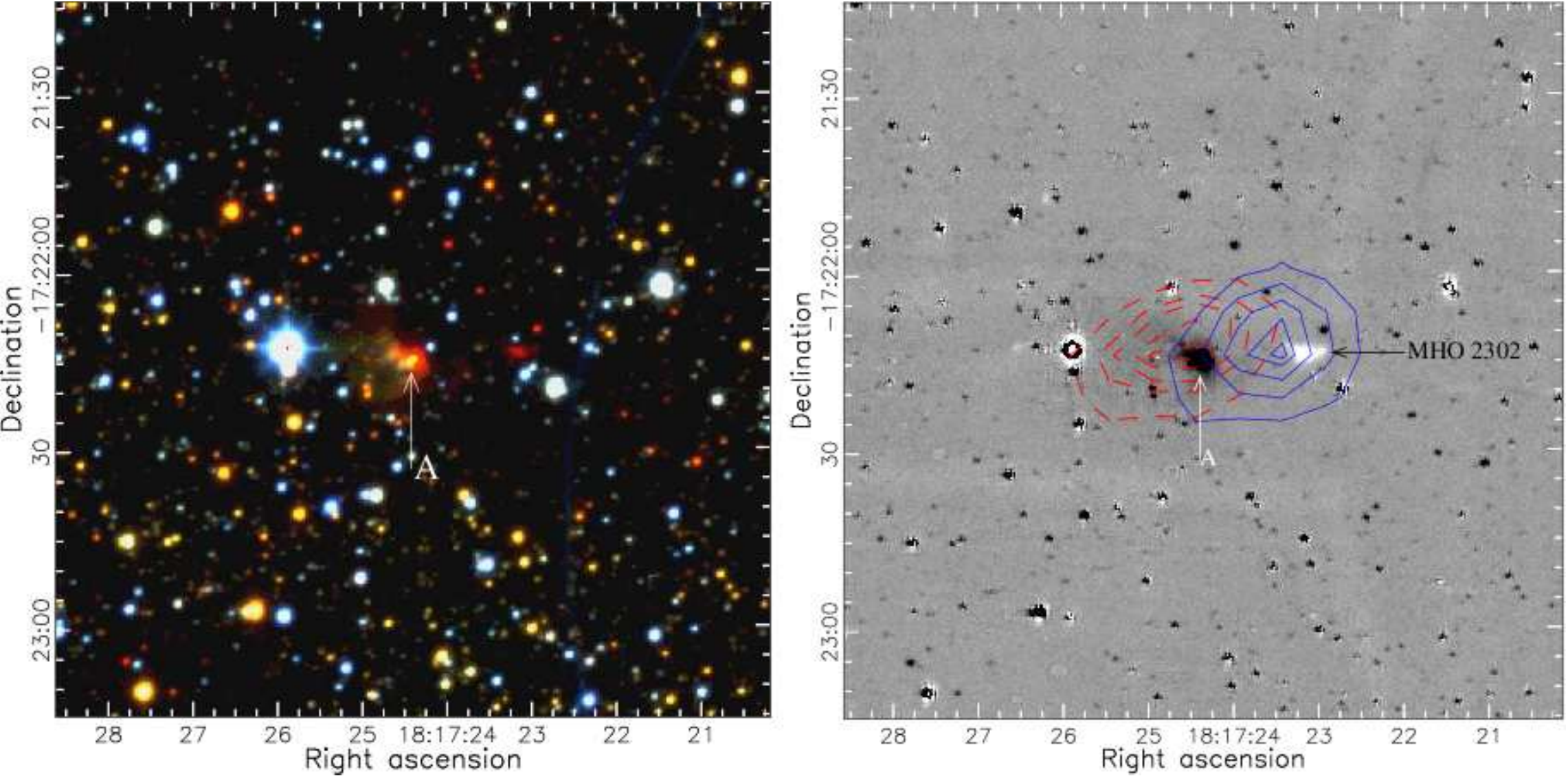}
\caption{Left: A $2\arcmin\times2\arcmin$ field of the $J$ (blue), 
$H$ (green), $H_2$ (red) colour composite image of IRAS~18144
observed using WFCAM. Right: The continuum-subtracted H$_2$ image 
of the same field. The blue (continuous) and
red (dashed) contours
are generated from the CO(3--2) intensity maps of the blue-
and red-shifted lobes of the outflow detected in our
JCMT+HARP observations, integrated in velocity ranges
of 23.3 -- 44.0\,km~s$^{-1}$ and 50.6 -- 71.3\,km~s$^{-1}$
respectively. The H$_2$ emission knot MHO~2302 and the 
outflow source candidate (`A') identified by \citet{Varricatt10}
are labelled.}
\label{Fig:JHH2_cs}
\end{figure*}

The observations were performed by dithering the
object to nine positions separated by a few arcseconds
and using a 2$\times$2 microstep, resulting in a pixel
scale of 0.2\arcsec~pixel$^{-1}$. The sky conditions
were clear.  Table \ref{tab:UKIRT_imaging_log} gives
a log of the WFCAM observations. Preliminary
reduction of the data was performed by the Cambridge
Astronomical Survey Unit (CASU). The photometric system
and calibration are described in \cite{Hewett06} and
\cite{Hodgkin09} respectively. The pipeline processing
and science archive are described in \cite{Irwin04} and
\cite{Hambly08}. Further reduction was carried out
using the Starlink packages {\sc{kappa}} and
{\sc{ccdpack}}.  The left panel of Fig. \ref{Fig:JHH2_cs} shows a
{\em J,H,H$_2$} ($J$-blue, $H$-green, $H_2$-red) colour
composite image in a 2$\arcmin\times$2$\arcmin$ field
centred on IRAS~18144. The long-integration $J$ and
$H$ images obtained on UT 20140620 and 20140621 and both
$H_2$ images obtained on 20140620 were averaged for
constructing Fig. \ref{Fig:JHH2_cs}. The narrow-band H$_2$
and [Fe {\sc{ii}}] images were continuum subtracted using
scaled {\em K} and {\em H} images respectively observed
closest in time as well as with good agreement in seeing,
adopting the procedure given in \citet{Varricatt10}.
The right panel of Fig. \ref{Fig:JHH2_cs}
shows the continuum-subtracted H$_2$ image. 
The extended emission features are continuum subtracted
well.  The point sources show positive or negative residuals
due to the difference in seeing between the two images or due
to reddening.
The H$_2$
emission knot (MHO~2302) and the outflow source candidate 
(`A') identified by \citet{Varricatt10} are labelled.
No line emission was detected in the
[Fe {\sc{ii}}] image, so it is not shown here. 

Source `A' of \citet{Varricatt10} is detected well in our 
$K$-band image as a highly reddened object embedded in 
nebulosity (see \S{\ref{cs_nc}}). Its detection is 
marginal in $H$, and most of the emission is from the 
nebulosity. It is not detected in $J$; only the nebulosity is seen.

\subsection[]{$L'$ and $M'$ imaging using UIST}

$L'$ (3.77\,$\mu$m) and $M'$ (4.69\,$\mu$m) imaging 
observations were obtained using the UKIRT 1--5\,$\mu$m 
Imager Spectrometer (UIST, \citet{Ramsay04}). UIST employs 
a 1024$\times$1024 InSb array. The 0.12\arcsec~pixel$^{-1}$
camera of UIST was used, which has an imaging field of 2\arcmin$\times$2\arcmin\
per frame.
The observations were performed by dithering 
the object to four positions on the array, separated by 
20\arcsec\ each in RA and Dec from the base position. Each pair 
of observations was flat-fielded and mutually subtracted, 
resulting in a positive and negative beam.  The final mosaic
was constructed by combining the positive and negative beams.
Table \ref{tab:UKIRT_imaging_log} gives the details of the
observations.

\begin{figure}
\centering
\includegraphics[width=8.4cm,clip]{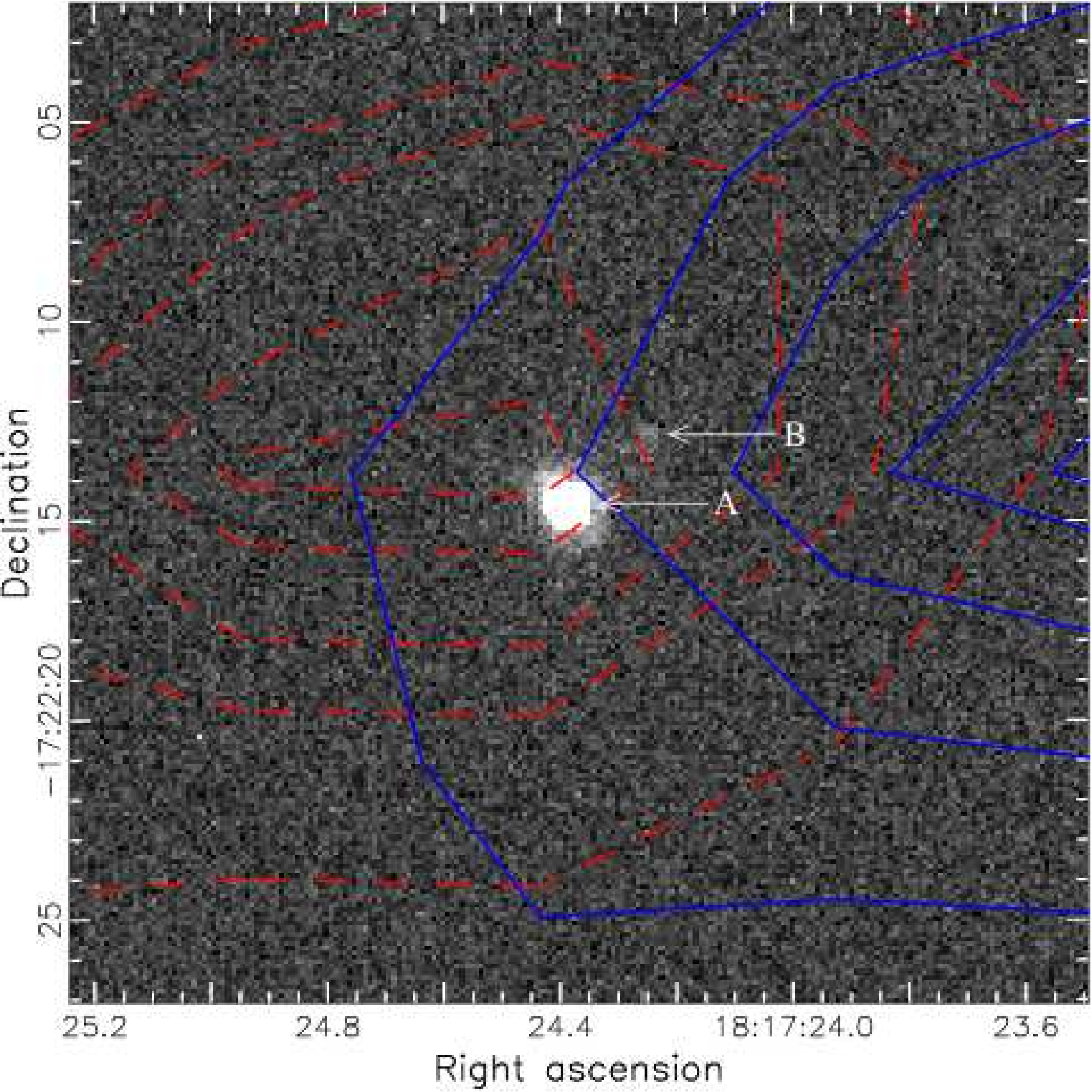}
\caption{A $25\arcsec\times25\arcsec$ field of the $M'$ 
image of IRAS~18144 observed using UKIRT and UIST. The 
blue (continuous) and red (dashed) contours are generated
from the integrated CO(3--2) maps of the red- and
blue-shifted lobes of the outflow. 
 }
\label{Fig:18144_UIST}
\end{figure}

The data reduction was carried out using the facility pipeline
{\textsc{oracdr}} \citep{cavanagh08} and using {\sc{kappa}} and
{\sc{ccdpack}}. Astrometric calibration was done using the
2MASS positions of the objects detected in our images. The UKIRT 
standard star GL748 was observed with the same dither pattern 
for photometric calibration. 
Fig. \ref{Fig:18144_UIST} shows a 25\arcsec$\times25\arcsec$
section of the $M'$ image. Source `A' was detected well in
$L'$ band as a single object. In $M'$, we detect a deeply embedded
source $\sim$2.6\arcsec\ NE of `A'; it is labelled `B' on the
figure. The magnitudes for sources `A' and `B' derived from
our images are given in Table \ref{tab:UIST_Mich_flx}.

\begin{table}
\caption{Results from UIST imaging}
\label{tab:UIST_Mich_flx}      
\centering               
\begin{tabular}{lllll}   
\hline\hline                                                            \\[-2mm]
Src  &RA      	&Dec		&\multicolumn{2}{c}{magnitudes$^{\mathrm{b}}$}  \\
ID      &                &                              &$L'$       &$M'$                \\
\hline                                                                                   \\[-2mm]
A       &18:17:24.375   &-17:22:14.71  &6.54 (0.04)  &5.57 (0.07)   \\
B       &18:17:24.239$^{\mathrm{a}}$   &-17:22:12.87$^{\mathrm{a}}$                   &Not det.   &9.57 (0.14)    \\
\hline
\end{tabular}
\begin{list}{}{}
\item $^{\mathrm{a}}$Derived from the Michelle
image after matching the coordinates of source `A' derived from
the $K$-band image; $^{\mathrm{b}}$A 4\arcsec-diameter photometric
aperture was used. `B' being faint, a 1.2\arcsec-diameter 
aperture was used and aperture correction was applied. The values given in
parenthesis are the 1-$\sigma$ errors in photometry.
\end{list}
\end{table}

\subsection[]{Michelle imaging}

Michelle \citep{Glasse97} is a mid-IR imager/spectrometer 
at UKIRT, employing an SBRC Si:As 320$\times$240-pixel array.  
It has a field of view of 67.2\arcsec$\times$50.4\arcsec 
with an image scale of 0.21\arcsec~pix$^{-1}$. We observed 
IRAS~18144 using Michelle on multiple nights in four filters 
centered at 7.9, 11.6, 12.5 and 18.5\,$\mu$m. The 
7.9\,$\mu$m filter has a 10\% passband, and the other filters 
have 9\% passbands.

\begin{figure}
\centering
\includegraphics[width=8.4cm,clip]{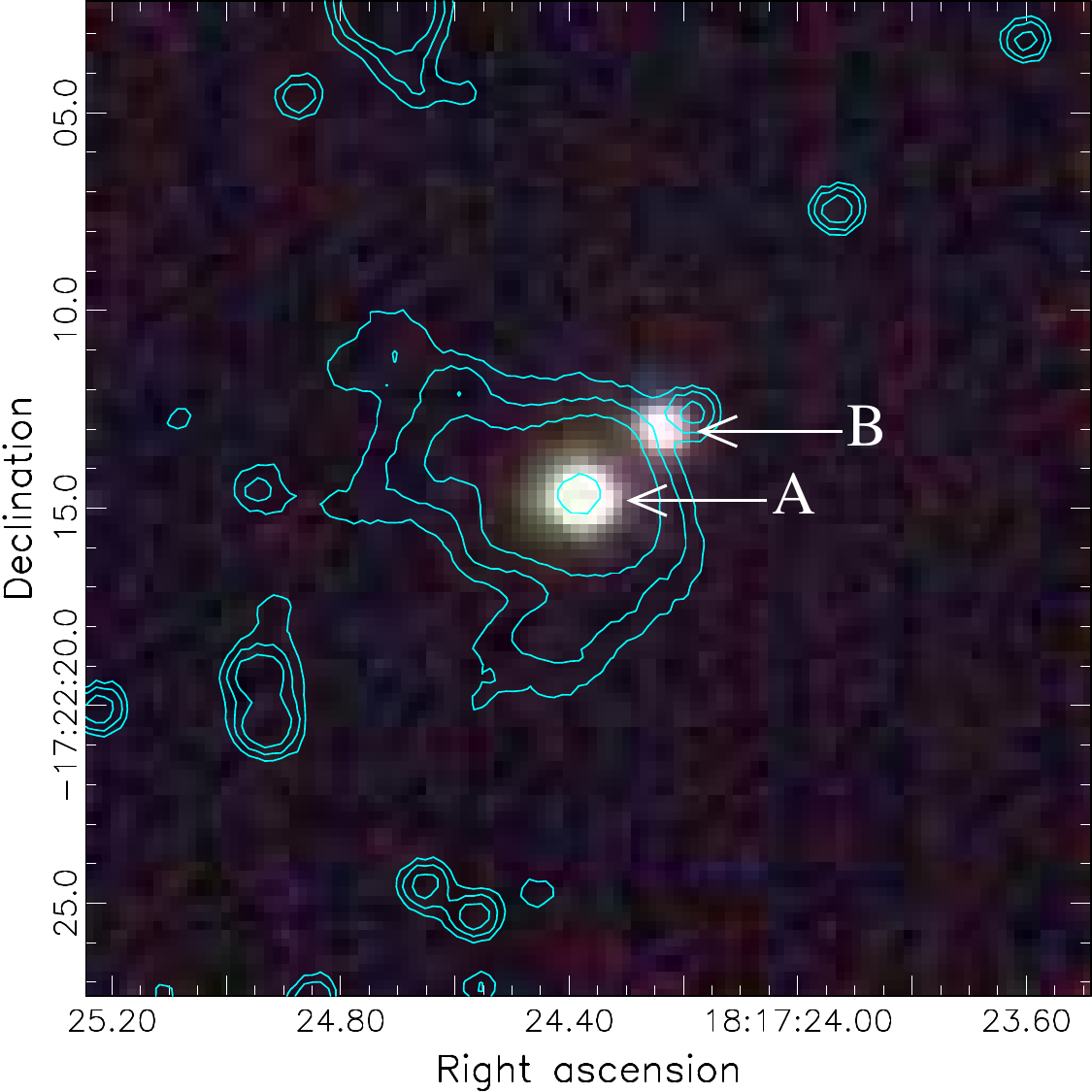}
\caption{
A colour composite
image (18.5\,$\mu$m (red), 12.5\,$\mu$m (green), 7.9\,$\mu$m (blue))
of a $25\arcsec\times25\arcsec$ field from our Michelle images.
The contours are generated from the
$K$-band image presented in \citet{Varricatt10}, and are at arbitrary
levels.
}
\label{Fig:18144_Mich}
\end{figure}

The sky conditions were photometric with the atmospheric 
opacity at 225\,GHz measured with the CSO (Caltech Submillimeter Observatory) dipper ($\tau_{225GHz}$)
was $\sim$0.07.
BS~6705 and BS~7525 were used as the standard stars 
on 20040329~UT, and BS~6869 was the standard for the 
rest of the observations.
Table \ref{tab:Mich_imaging_log} shows the details of 
the observations. The details of the data reduction are 
the same as in \citet{Varricatt13}.
Astrometric corrections were applied by adopting the position 
of source `A' from \citet{Varricatt10}.
Fig. \ref{Fig:18144_Mich} shows a
colour composite image in a
25\arcsec$\times$25\arcsec\ region constructed from 
our Michelle images at 18.5, 12.5 and 7.9\,$\mu$m.
Aperture photometry was performed using an aperture
of diameter 14 pixels (2.94$\arcsec$).
The flux densities measured for sources `A' and `B' are given in
Table \ref{tab:Mich_imaging_log}. 

\begin{table*}
\centering
\begin{minipage}{140mm}
\caption{Observations using UKIRT and Michelle}
\label{tab:Mich_imaging_log}
\begin{tabular}{@{}llllllll@{}}
\hline
UTDate    & Filter  & Exp. time &Coadds  &Int. time &FWHM      &\multicolumn{2}{c}{Flux density in Jy$^{\mathrm{a}}$}\\
          &         & (sec)     &   &(sec)     &(arcsec)       &Source `A'&Source `B'\\
\hline
20151107  & 7.9\,$\mu$m  &0.08     &155    &148.8               &0.73   &2.17 (0.16) &0.85 (0.1) \\
20160826  & 7.9\,$\mu$m  &0.08     &155    &248.0               &0.72   &2.02 (0.04) &0.92 (0.1) \\[3mm]
20151107  & 11.6\,$\mu$m &0.09     &144    &155.52              &0.75   &2.23 (0.04) &0.40 (0.05)\\
20160826  & 11.6\,$\mu$m &0.09     &144    &259.2               &0.76   &1.93 (0.04) &0.40 (0.12)\\[3mm]
20040329  & 12.5\,$\mu$m &0.08	   &155    &99.2		&0.82   &3.16 (0.18) &0.96 (0.13)\\
20160826  & 12.5\,$\mu$m &0.07     &168    &141.12              &0.83   &2.55 (0.06) &0.93 (0.06)\\
20160920  & 12.5\,$\mu$m &0.07     &168    &161.28              &0.87   &2.74 (0.13) &0.93 (0.07)\\[3mm]
20151107  & 18.5\,$\mu$m &0.03     &338    &162.24              &1.08   &5.27 (0.09) &2.41 (0.13)\\
20160826  & 18.5\,$\mu$m &0.03     &338    &283.92              &1.08   &4.54 (0.04) &2.25 (0.06)\\
20160920  & 18.5\,$\mu$m &0.03     &338    &162.24              &1.02   &5.63 (0.36) &2.28 (0.24)\\
\hline
\end{tabular}
\begin{list}{}{}
\item $^{\mathrm{a}}$The values given in
parenthesis are the 1-$\sigma$ errors in the flux density measured in the four beams from the Michelle
image observed with chopping and nodding.
\end{list}
\end{minipage}
\end{table*}

\subsection{Archival imaging data}

The field containing IRAS~18144 was covered in the {\it Spitzer}
GLIMPSE survey using the Infrared Array Camera \citep[IRAC;][]{Fazio04},
and the MIPSGAL survey using MIPS \citep{Rieke04}. We downloaded 
the IRAC images and catalogue and the MIPS images from the {\it Spitzer}
Heritage Archive\footnote{http://sha.ipac.caltech.edu/applications/Spitzer/SHA/}. Source `A' was detected well in the IRAC bands 
1--4 (centered at 3.5, 4.5, 5.8 and 8.0\,$\mu$m respectively).  
`B' was not detected in bands 1 and 2, and was only 
marginally detected in band 3. It was detected well in  
band 4, but the psf merges with that of `A'.
The {\it Spitzer} catalog lists its location as
($\alpha$=18:17:24.25, $\delta$=-17:22:13.16), which agrees with
our position (Table \ref{tab:UIST_Mich_flx}) within the 0.3\arcsec\ error in the
coordinates given in the GLIMPSE catalogue.
IRAC 5.8 and 8.0\,$\mu$m flux densities 
listed for this source are 264.0$\pm$30.2 and 534.4$\pm$56.5\,mJy
respectively. However, note that in UIST $M'$ (4.7\,$\mu$m), where `B'
is resolved well from `A', source `B' has a flux density of only 24.2$\pm$5\,mJy (9.57 mag.;
Table \ref{tab:UIST_Mich_flx}).
The flux estimates 
of `B' are likely to be affected by the proximity of `A'.
Therefore, we do not use the 5.8\,$\mu$m flux of `B' in our
analysis. The source is detected in the MIPS band 1 (23.68\,$\mu$m)
images at ($\alpha$=18:17:24.24, $\delta$=-17:22:12.61), but it is 
saturated making the photometry unusable.  
At a 2.45\arcsec~pix$^{-1}$ spatial resolution of the MIPS image, `A' 
and `B' are not resolved, but it should be noted that the centroid 
is much closer to `B' than to `A'.

IRAS~18144 was observed in the WISE \citep{Wright10} and
AKARI \citep{Murakami07} surveys. The sources `A' and `B' are 
not resolved by WISE. Our imaging at higher angular resolution 
shows that the emission from `A' will the dominant source in 
the WISE bands W1, W2 and W3 at  3.4, 4.6, 12\,$\mu$m
respecitvely, so the magnitudes measured in those bands are 
not used in our analysis. Table \ref{tab:Mich_imaging_log} 
shows that `A' contributes $\sim$70\% of the total flux at 
18.5\,$\mu$m, so we use the band W4 (22\,$\mu$m) magnitude of
-1.191$\pm$0.011 as only an upper limit in the SED analysis.
The flux densities measured in the AKARI-IRC 9 and 18\,$\mu$m 
bands were not used in this study as they will be dominated by 
source `A', and our Michelle observations at better spatial
resolution cover these wavelengths.
AKARI-FIS 90 and 140\,$\mu$m flux densities were also not used
as they will be affected by the large
bandwidth of these filters. The FIS 65
and 160$\mu$m flux densities were used after applying 
the colour correction given in \citet{Yamamura10}.

We used the Herschel PACS \citep{Poglitsch10} (70 and 160\,$\mu$m), and SPIRE
\citep{Griffin10} (250\,$\mu$m, 350\,$\mu$m and 500\,$\mu$m) 
level\,2.5 maps (observation ID \#1342218999 and 
\#1342219000) publicly available from the Herschel Science Archive. 
`A' and `B' are not resolved in the Herschel maps.
Aperture photometry of the source was performed using HIPE and 
aperture corrections were applied. Table \ref{tab:Herschel_flx} 
gives the flux densities for sources `A' and `B' combined, 
measured from the PACS and SPIRE maps.

\begin{table}
\centering
\begin{minipage}{140mm}
\caption{Flux density of `A+B' from the Herschel maps}
\label{tab:Herschel_flx}
\begin{tabular}{@{}llll@{}}
\hline
Wavelength 	&Aperture	&Flux density	&Error\\
($\mu$m) 	&diameter (arcsec) & (Jy)	& (Jy) \\
\hline
70	&24  	&525.6		&20.5 \\
160	&44 	&938.6		&27.7  \\
250 	&44 	&454.5		&18.8 \\
350	&60 	&191.7		&12.4 \\
500 	&84 	&74.7		&2.2  \\
\hline
\end{tabular}
\end{minipage}
\end{table}

\subsection{Near-IR spectroscopy using UIST}

We obtained near-IR spectroscopic observations of IRAS~18144 
using UKIRT and UIST on UT\,20140824. The sky conditions were 
clear with good seeing ($\sim$0.24\arcsec\ in the $K$ band). The HK 
grism was used along with a 4-pix-wide (0.48\arcsec) slit, 
giving a wavelength coverage of 1.395--2.506\,$\mu$m, and spectral 
resolution of $\sim$500. Flat field observations were obtained 
prior to the target observations by exposing the array to a 
black body mounted on the instrument. For wavelength calibration, 
the array was exposed to an Argon arc lamp.
An early-type telluric standard was also observed at the same
airmass as that of the target. We adopted a slit angle of
85.2$^{\circ}$ West of North so that source `A' and the
H$_2$ emission feature MHO\,2302 are both on the slit.
For the target field, we nodded the telescope between the
source position and a blank field nearby to enable good sky
subtraction. The total on-chip exposure time for the target
was 600\,sec.

\begin{figure*}
\centering
\includegraphics[width=17.7cm,clip]{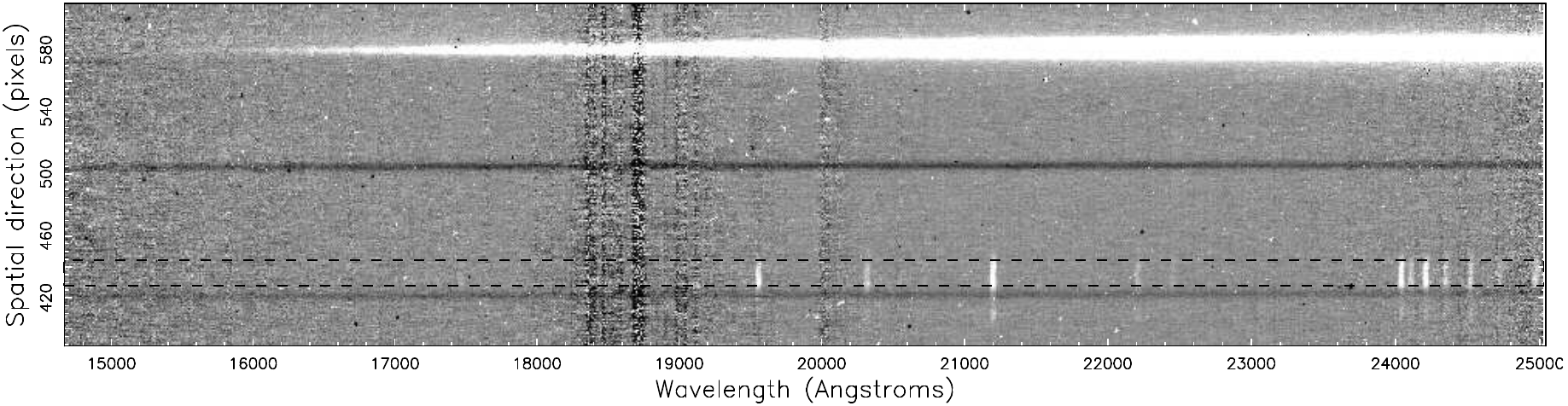}
\caption{UIST spectral image of IRAS~18144 containing 
source `A' and the outflow lobe MHO\,2302.
The image covers a wavelength regime of 14480--25059~\AA\ on 
the X axis and has a spatial extent of 221 pixels (26.5$\arcsec$) 
on the Y axis. The dashed box show the region within which 
the spectrum of MHO\,2302 is extracted
}
\label{Fig:UIST_spim}
\end{figure*}

\begin{figure*}
\centering
\includegraphics[width=17.7cm,clip]{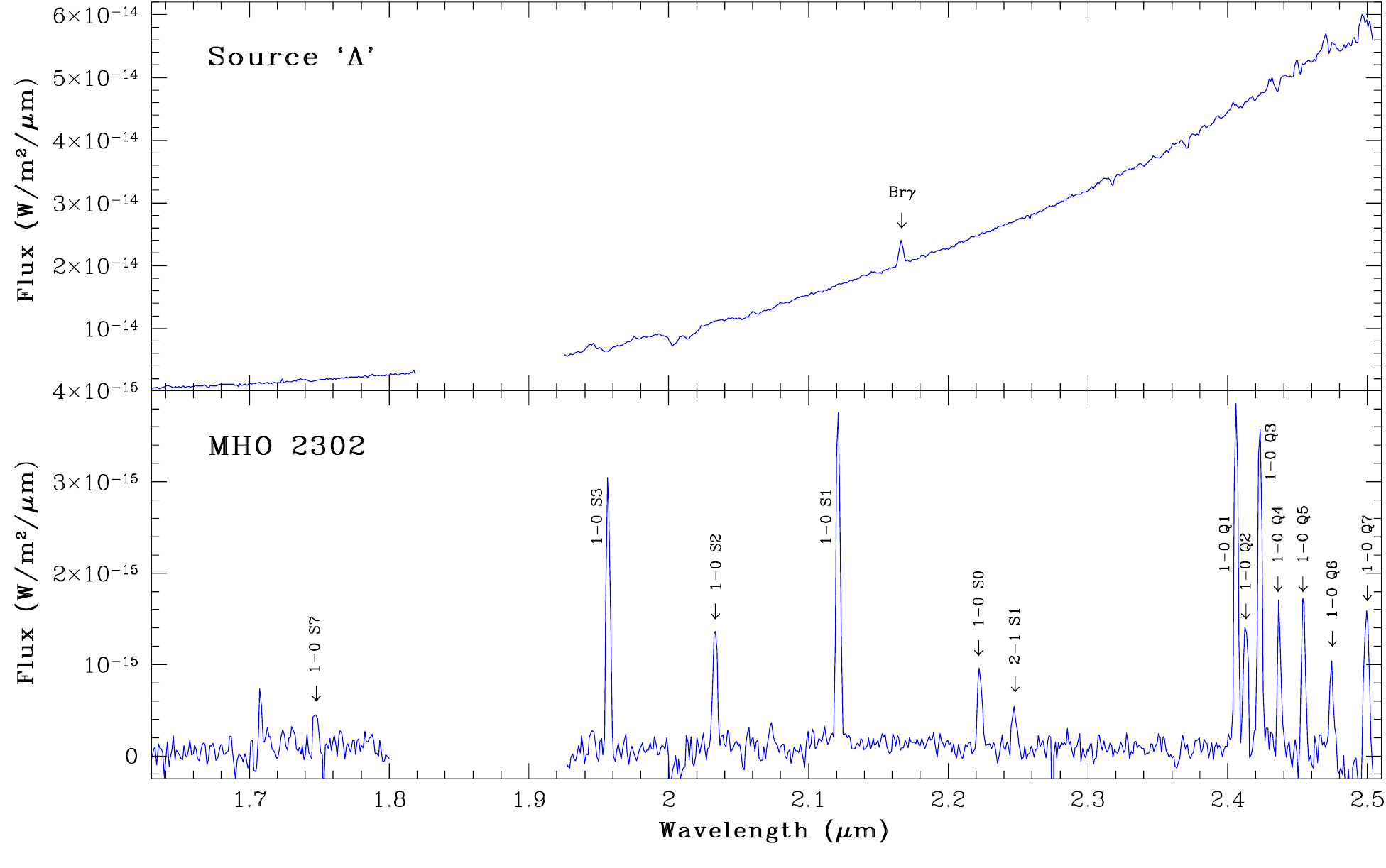}
\caption{The upper panel shows the spectrum of source `A'.
The lower panel shows the spectrum of MHO\,2302, integrated
over a field of 7.68\,arcsec$^2$ (the region enclosed by the
dashed box drawn in Fig. \ref{Fig:UIST_spim}).
}
\label{Fig:UIST_sp_csjet}
\end{figure*}

The data reduction was carried out using the UKIRT pipeline
{\sc{oracdr}} and Starlink {\sc{kappa}} and {\sc{figaro}} 
\citep{currie08}. 
The spectra from the two nodded beams of the reduced spectral
image of standard star were 
extracted, averaged and wavelength calibrated. It was then 
divided by the spectrum of a blackbody of temperature similar 
to its photosperic temperature, and the photospheric absorption 
lines were interpolated out. The flat-fielded and sky-subtracted 
target spectal image was wavelength calibrated and
divided by the interpolated standard star spectrum.
The ratioed spectrum was 
then flux calibrated using the magnitudes of source `A' derived 
from our photometry. Fig. \ref{Fig:UIST_spim} shows the calibrated 
spectral image of IRAS~18144. The extracted spectrum of source `A' 
is shown in the upper panel of Fig. \ref{Fig:UIST_sp_csjet}.
The most prominent feature in the spectrum of source `A' is 
the Br$\gamma$ emission line. 
The emission from MHO\,2302 is composed purely of the ro-vibration 
lines of H$_2$. The spectrum of MHO\,2302 extracted in 16 rows 
(7.68\,arcsec$^2$; the region shows in the dashed box in 
Fig. \ref{Fig:UIST_spim}) is shown in the lower panel of
Fig. \ref{Fig:UIST_sp_csjet}. The integrated intensities of the 
H$_2$ lines measured from this spectrum are given in 
Table \ref{tab:H2_lines}.

\begin{table}
\caption{H$_2$ line intensities}
\label{tab:H2_lines}      
\centering                      
\begin{tabular}{lllllll}        
\hline\hline                                    \\[-2mm]
$\lambda_{air}$& Species&Intensity &Error    &Einstein &g$^{\mathrm{a}}$ &E($\upsilon$,J)   \\
(\AA)  &     & W/m$^2$   &W/m$^2$  &A (s$^{-1}$)  &   &(K)      \\
\hline
22235& 1--0 S0&  3.51E-18 &  4.00E-20  &2.53E-7 &     5  & 6470   \\
21218& 1--0 S1&  1.34E-17 &  2.00E-19  &3.47E-7 &     21 & 6960   \\
20338& 1--0 S2&  5.10E-18 &  5.00E-20  &3.98E-7 &     9  & 7580   \\
19576& 1--0 S3&  1.05E-17 &  2.00E-19  &4.21E-7 &     33 & 8370   \\
17480& 1--0 S7&  1.65E-18 &  1.50E-19  &2.98E-7 &     57 & 12800  \\
22477& 2--1 S1&  2.06E-18 &  4.00E-20  &4.98E-7 &     21 & 12600  \\
24066& 1--0 Q1&  1.27E-17 &  2.00E-19  &4.29E-7 &     9  & 6150   \\
24134& 1--0 Q2&  5.40E-18 &  2.00E-19  &3.03E-7 &     5  & 6470   \\
24237& 1--0 Q3&  1.24E-17 &  3.00E-19  &2.78E-7 &     21 & 6960   \\
24375& 1--0 Q4&  4.78E-18 &  1.00E-19  &2.65E-7 &     9  & 7580   \\
24548& 1--0 Q5&  6.43E-18 &  2.00E-19  &2.55E-7 &     33 & 8370   \\
24756& 1--0 Q6&  4.21E-18 &  1.50E-19  &2.45E-7 &     13 & 9290   \\
25001& 1--0 Q7&  8.88E-18 &  5.00E-19  &2.34E-7 &     45 & 10300  \\
\hline
\end{tabular}
\begin{list}{}{}
\item $^{\mathrm{a}}$ Statistical weight
\end{list}
\end{table}

\subsection{Sub-mm continuum observation using SCUBA-2}

We observed IRAS~18144 using the James Clerk Maxwell Telescope 
(JCMT) and Submillimetre Common-User Bolometer Array 2 (SCUBA-2;
\cite{Holland13}) on UT~20140626. SCUBA-2 maps simulteneously 
at 450 and 850\,$\mu$m. Each band uses four 32$\times$40 Transition Edge 
Sensor (TES) arrays, with a main beam
size of 7.9\arcsec\ and 13.0\arcsec\ respectively at 450\,$\mu$m
and 850\,$\mu$m. 

The observation was performed by moving the telescope in a
``CV Daisy'' pseudo-circular 
pattern\footnote{See http://www.eaobservatory.org/jcmt/instrumentation/\\continuum/scuba-2/
for more details about observatios using SCUBA-2}.
A field of diameter of $\sim$12\arcmin\ was
obtained. The weather conditions 
were good with $\tau_{225\,GHz}$=0.08 during the observations. 
Pointing and extinction corrections were applied using pointing 
and flux standards observed before the target observations. 
The data reduction was performed using the Starlink package
{\sc{smurf}}\footnote{Also see
http://www.starlink.ac.uk/docs/sc21.htx/sc21.html}.
The maps were sampled down to 2\arcsec~pix$^{-1}$ 
at 450\,$\mu$m and 4\arcsec~pix$^{-1}$ at 850\,$\mu$m. We 
reach RMS noise levels of 2.04\,mJy\,arcsec$^{-2}$ 
(213\,mJy/beam) and 0.031\,mJy\,arcsec$^{-2}$ (7\,mJy/beam) 
respectively at the centers 
of the 450\,$\mu$m and 850\,$\mu$m maps.

\begin{figure}
\centering
\includegraphics[width=8.4cm,clip]{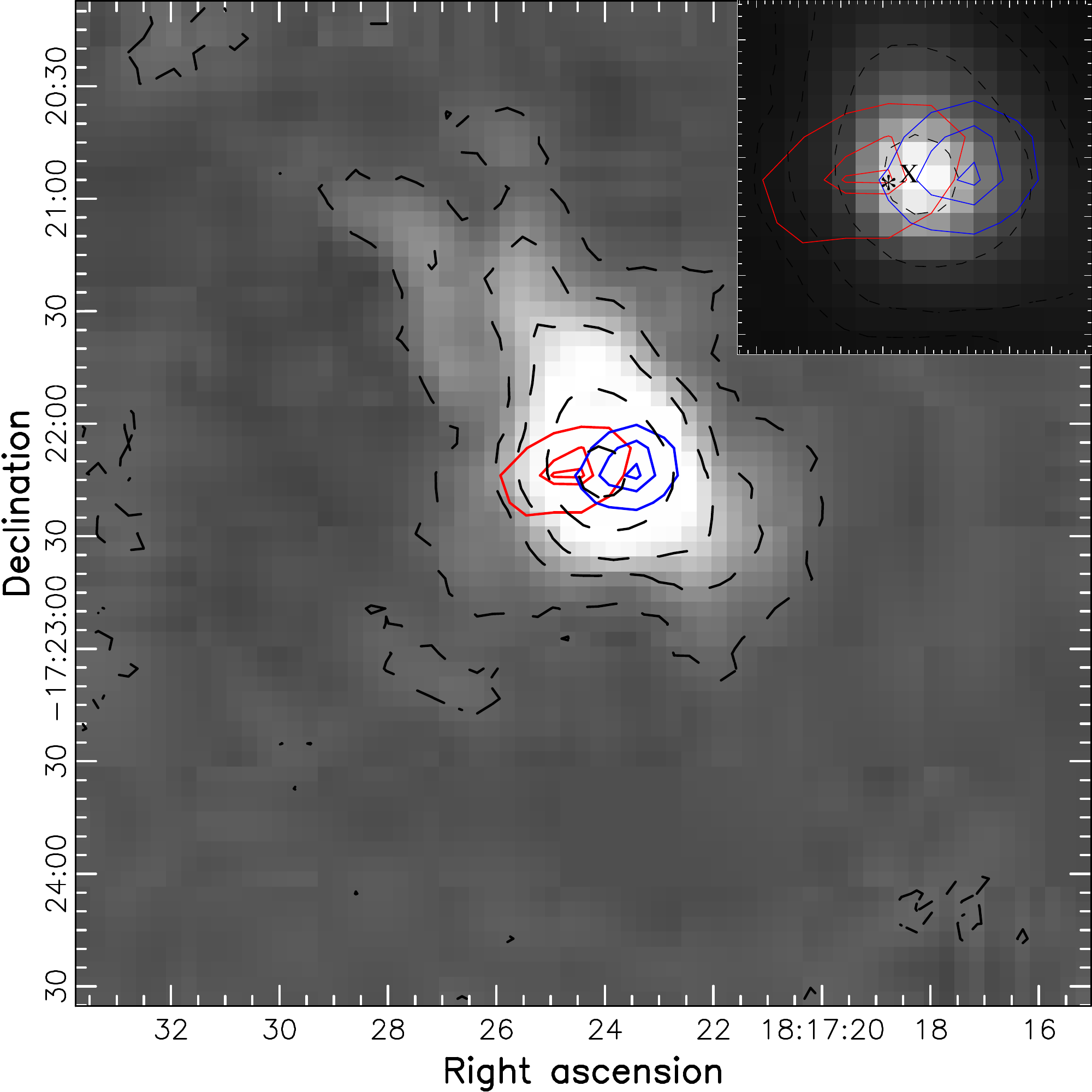}
\caption{A 4.5\arcmin$\times$4.5\arcmin field extracted
from the SCUBA2 850$\mu$m map.  The dashed black contours
show flux densities of 12, 3, 1, 0.5 and 0.1\,mJy~arcsec$^{-2}$ 
respectively on the 850$\mu$m map.  The blue and red contours 
are generated from the CO(3--2) intensity maps of the blue- 
and red-shifted lobes of the outflow detected in our 
JCMT+HARP observations, integrated in velocity ranges of 
23.3 -- 44.0\,km~s$^{-1}$ and 50.6 -- 71.3\,km~s$^{-1}$ 
respectively. The blue contours are at 51, 40 and 
25\,K\,km\,s$^{-1}$ and the red contours are at 34.5, 
30 and 15\,K\,km\,s$^{-1}$. A 1\arcmin$\times$1\arcmin 
image of the central source is shown in high contrast in the inset. 
`$\ast$' shows the location of source `A' and `$\times$' shows the
location of source `B'.}
\label{Fig:SCUBA2_850_CO}
\end{figure}

Fig. \ref{Fig:SCUBA2_850_CO} shows the central 
4.5\arcmin$\times$4.5\arcmin\ field extracted from the 
850-$\mu$m SCUBA2 map. The SCUBA2 maps reveal a dense
core associated with IRAS~18144. We measure an FWHM of 18.5\arcsec\ and 22.6\arcsec\
respectively at 450\,$\mu$m and 850\,$\mu$m. These are larger
than the average FWHM of 9.35\arcsec\ and 14.2\arcsec\
respectively at 450\,$\mu$m and 850\,$\mu$m for two point
source standards observed prior to the target observation.
The centroid of the core is at $\alpha$=18:17:24.01, $\delta$=-17:22:12.05. 
This location is only 3.4\arcsec\ from source `B' and is 
closer to source `B' than to `A'. Also plotted on the 
figure are the contours of the red- and blue-shifted 
lobes of the CO outflow (see \S\,\ref{COdata}), 
which appear to be centered on the sub-mm source. 
The maps also reveal filamentary emission extending in the 
NE and SW of the central source, more prominent towards 
the NE. The longest filament seen in our maps extends to a distance 
of 2.5\arcmin\ from the central source, which is 3.15\,pc.
Table \ref{tab:SCUBA2_flx} shows the flux
density measured in different circular apertures around the 
point source. We measure a peak flux density of 
31\,Jy\,beam$^{-1}$ and 4\,Jy\,beam$^{-1}$ at 450 and 850\,$\mu$m
respectively on the point source.

\begin{table}
\caption{Flux density of `A+B' measured in different circular apertures
from our SCUBA2 maps}
\label{tab:SCUBA2_flx}
\begin{tabular}{@{}lll@{}}
\hline
Aperture diameter &  \multicolumn{2}{c}Flux density (Jy)\\
(arcsec) & 450\,$\mu$m &  850\,$\mu$m \\
\hline
30  & 66.6  & 5.8  \\
40  & 82.2  & 7.4  \\
45  & 87.9  & 8.0  \\
60  & 99.6  & 9.3  \\
80  & 109.4 & 10.4 \\
160 & 120.1 & 11.8 \\
\hline
\end{tabular}
\end{table}

\subsection{Heterodyne observations using HARP}

\subsubsection {CO\,($J$=3--2) data}
\label{COdata}

We obtained heterodyne mapping observations in CO with the JCMT on 20140731~UT.
The observations were performed using HARP \citep{buckle09} as the front end 
and the ASCIS autocorrelator as the back end, in position-switched raster-scan 
mode with quarter array spacing.
$^{12}$CO\,(3--2) (345.796\,GHz) and H$^{13}$CN (345.3398\,GHz),
and $^{13}$CO\,(3--2) (330.588\,GHz) and
C$^{18}$O~(3--2) (329.331\,GHz) were simultaneously observed
in two different dual sub-band settings of ASCIS.
This gives a bandwidth of 250\,MHz and
a resolution of 61\,kHz (0.055\,km~s$^{-1}$) for each line. The
scans were performed in a 3$\arcmin\times$3$\arcmin$ area.
The CSO $\tau_{225\,GHz}$ was
0.12 for $^{12}$CO/H$^{13}$CN observations, and 0.128 for $^{13}$CO/C$^{18}$O
observations.  
The pointing accuracy on source
{is better than 2\arcsec.}

The data were reduced using the {\sc oracdr} pipeline, 
which performs a quality-assurance check on the time-series data,
followed by trimming and de-spiking, and an iterative
baseline removal routine before creating the final group files. 
After binning to a resolution of 0.423\,km~s$^{-1}$ (please see the
next paragraph), we achieve an RMS noise level of 0.32, 0.31, 0.81 and 1.27\,K in
$^{12}$CO, H$^{13}$CN, $^{13}$CO and C$^{18}$O respectively.

We also downloaded JCMT archival data in $^{12}$CO\,(3--2),
$^{13}$CO\,(3--2) and C$^{18}$O~(3--2) lines. The $^{12}$CO
observations were performed on 20080419~UT with a 1000\,MHz band
width 2048 channel set up giving a velocity resolution of
0.423\,km~s$^{-1}$. The CSO $\tau_{225\,GHz}$ was 0.115.
The $^{13}$CO and C$^{18}$O lines
were observed on 20080706~UT in a single setting with 250\,MHz band
width 4096 channels for each line, giving a velocity resolution of
0.055 and 0.056\,km~s$^{-1}$ respectively.
Similar to our observations, the archival observations were also
obtained in position-switched raster scan mode, albeit with half
array spacing. The CSO $\tau_{225\,GHz}$ was 0.09 for the 
$^{13}$CO/C$^{18}$O observations. 
These spectra were binned
to a 0.423\,km~s$^{-1}$ resolution as in the $^{12}$CO line.
The $^{12}$CO, $^{13}$CO and C$^{18}$O archival spectra 
reached an RMS noise level of 0.41, 0.56 and 0.88\,K respectively.

We used our new $^{12}$CO observations 
for the figures presented in this paper.
Column densities were estimated separately from our own observations
and the archival observations to show the repeatability, and the
average was used in the discussions.
The results from the analysis of the CO data are presented in
\S\ref{CO-outflow}.  No H$^{13}$CN line emission was detected in our
observation.

\subsubsection{HCO$^{+}$\,(4-3)}
\label{HCO+}
\begin{figure}
\centering
\includegraphics[width=8.4cm,clip]{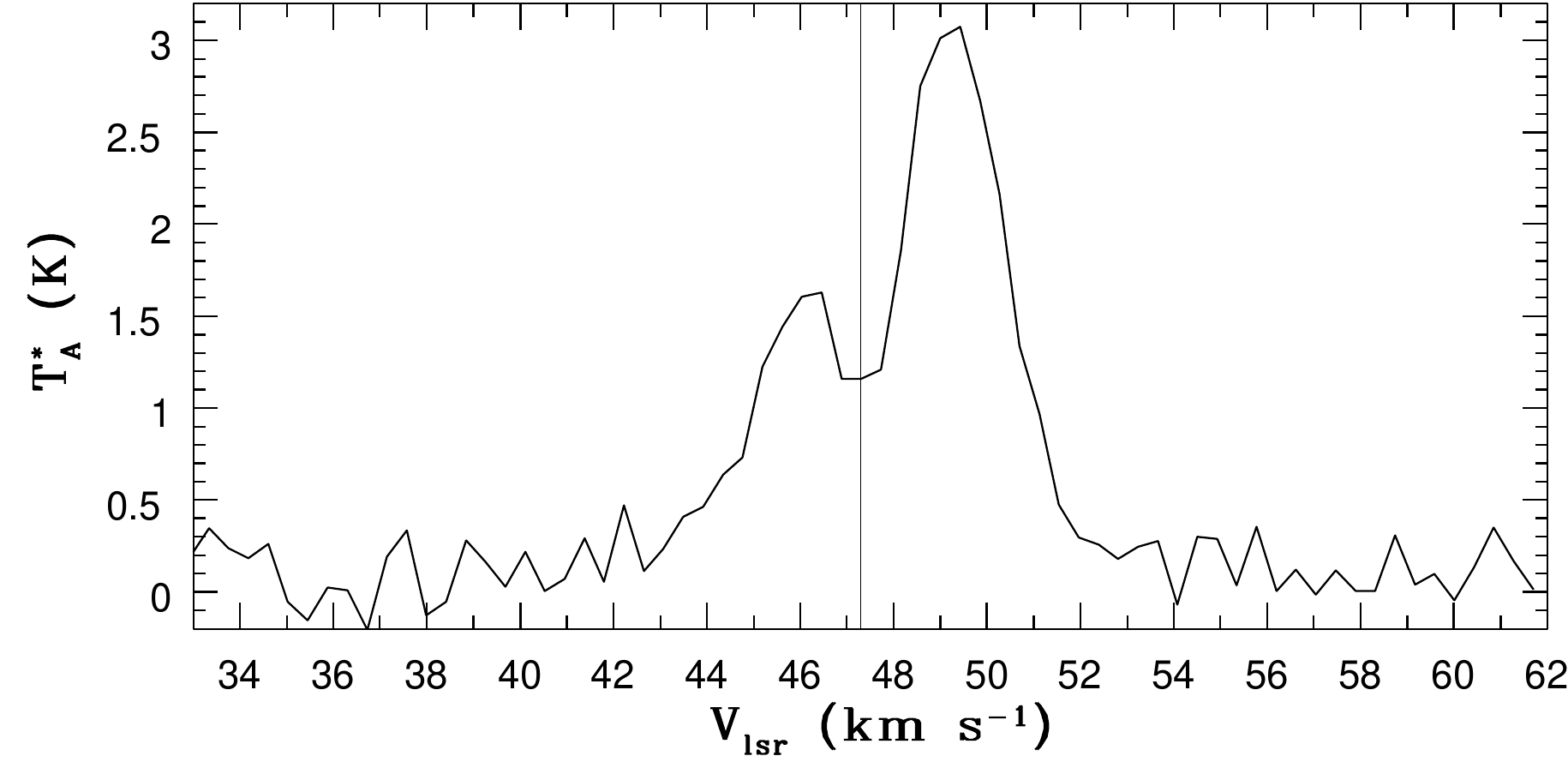}
\caption{The HCO$^{+}$ line at the peak,
binned down to a velocity
resolution of 0.42~km~s$^{-1}$.
The vertical line shows the systemic velocity of 47.3~km~s$^{-1}$.
The peak T$_{mb}$ observed was 5.17\,K (T$_A^*$=3.31)}
\label{Fig_IRAS18144_HCO}
\end{figure}

Archival data of IRAS~18144 in the HCO$^+$\,(4-3) line (356.7\,GHz)
were downloaded from the JCMT science archive at CADC.
The observations were performed on 20100405\,UT with HARP 
in jiggle map mode covering a field
of 2\arcmin$\times$2\arcmin. 
A 256\,MHz 4096 channel
HARP setup was used giving a velocity resolution of 0.051\,km~s$^{-1}$.
The mean atmospheric opacity at 225\,GHz, measured with the CSO
dipper, was 0.084 for the period of the observations.  The archival
data was reduced using {\sc{oracdr}}, further reduction
was carried out using {\sc{kappa}}.
After binning the velocity resolution down to 0.423\,km~s$^{-1}$,
we get an RMS of 0.11\,K. Fig. \ref{Fig_IRAS18144_HCO} shows
the spectrum (at the peak) after binning.

\section{Results and discussions}

\subsection{The outflow}

\subsubsection{From the sub-mm data}
\label{CO-outflow}

\begin{figure*}
\centering
\includegraphics[height=12cm,clip]{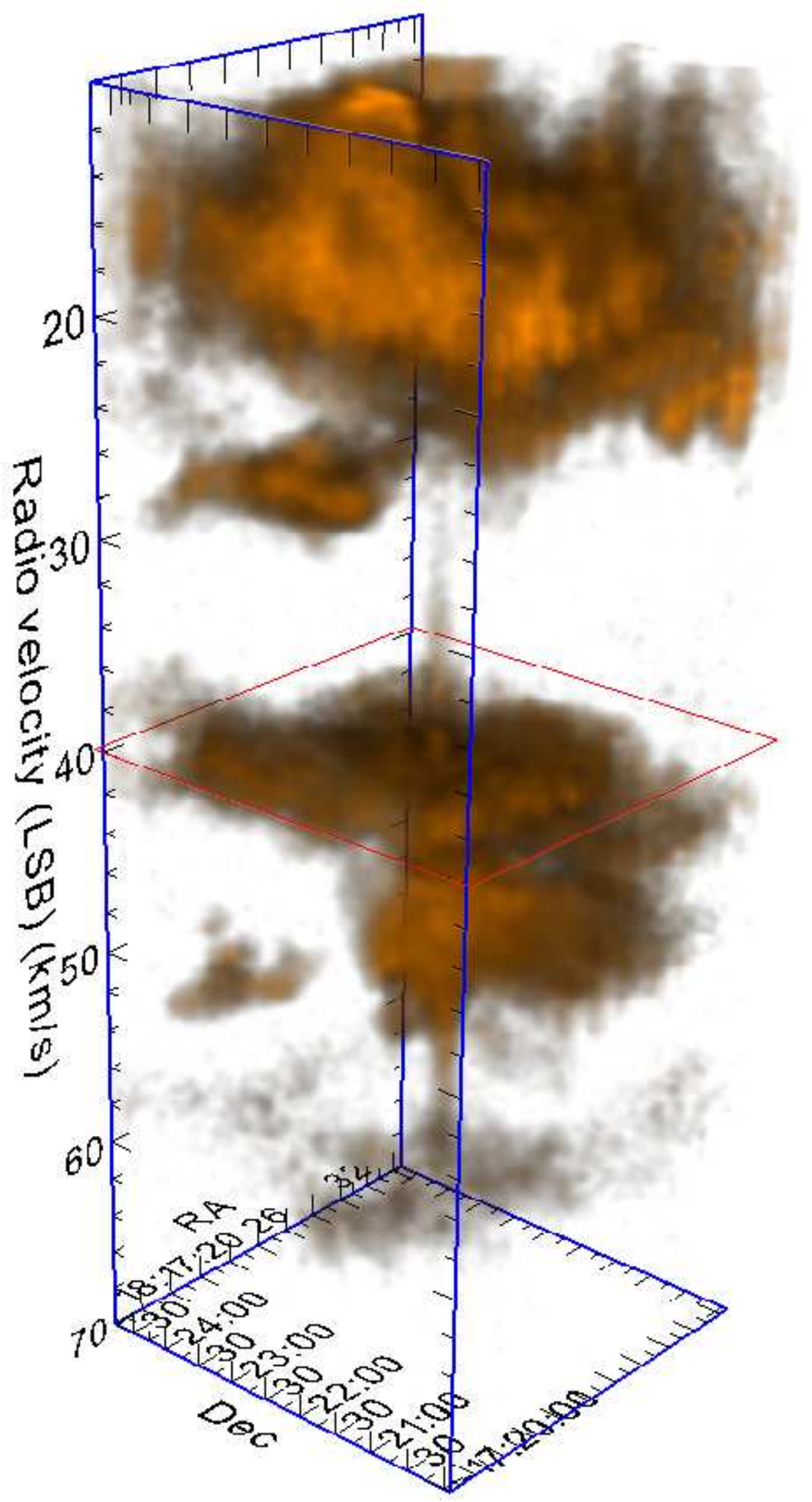}
\includegraphics[height=12cm,clip]{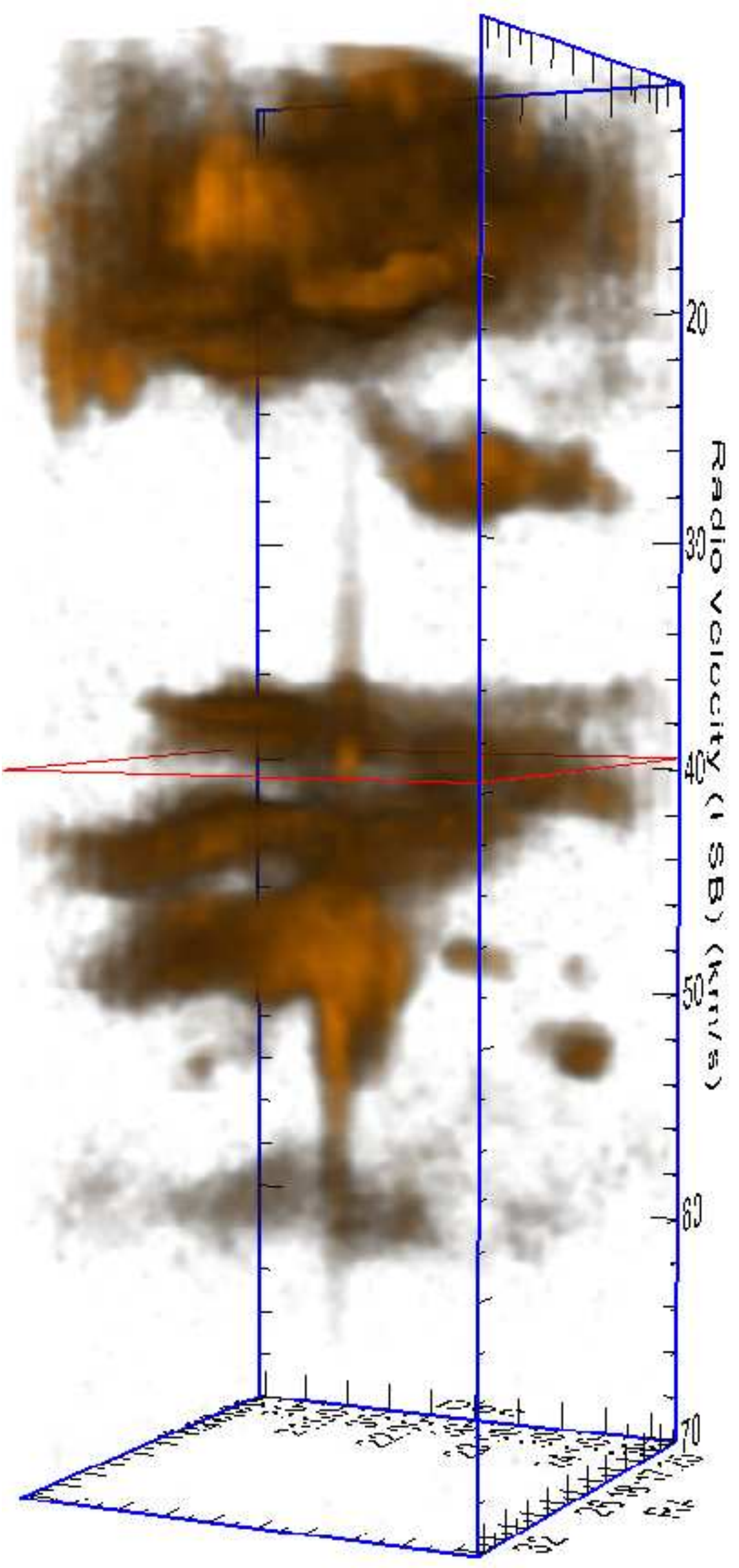}
\includegraphics[height=12cm,clip]{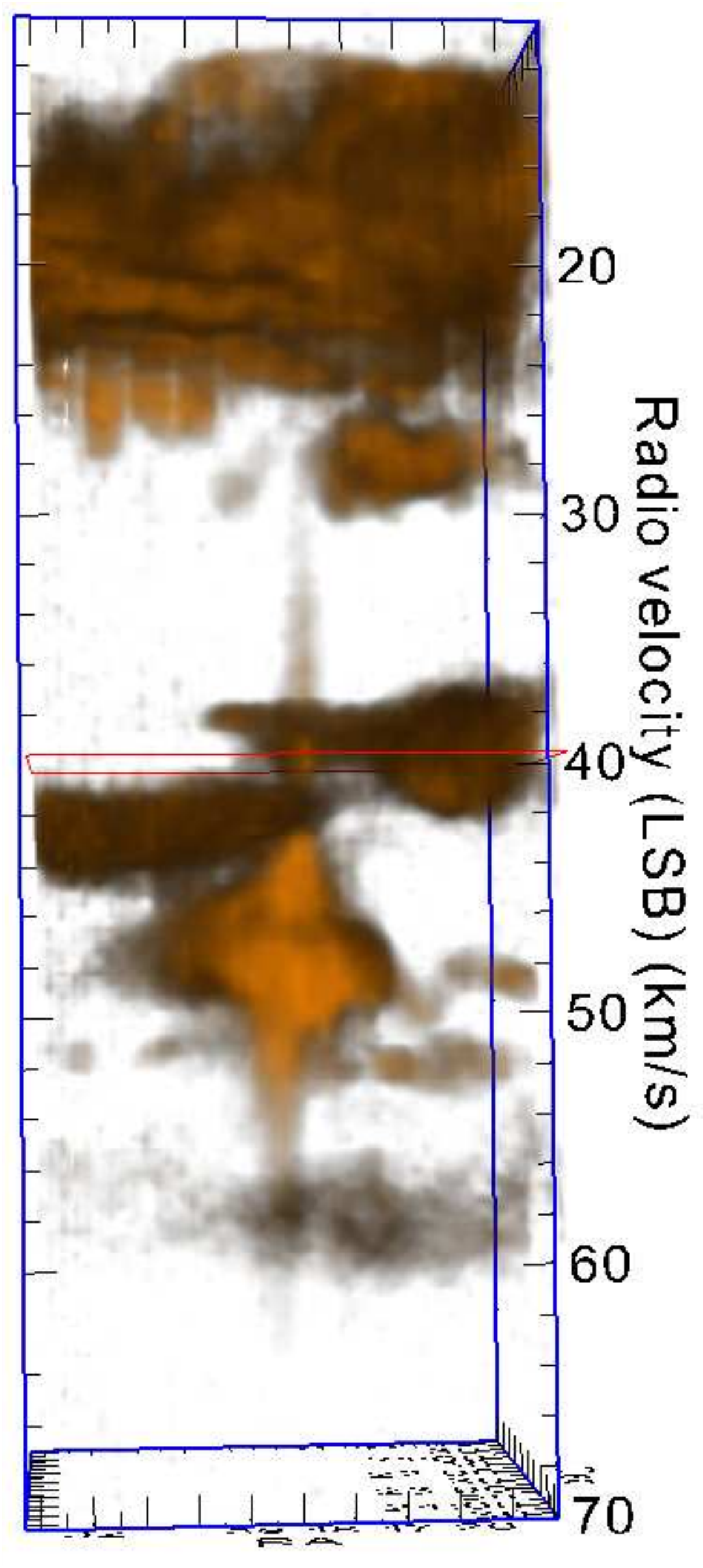}
\caption{Three different orientations of the $^{12}$CO (3--2) spectral cube
obtained on 20140731~UT showing the outflow, and the emission
from the ambient and line-of-sight regions. The images cover
a field of 5\arcmin$\times$5\arcmin\ and a radial velocity range of 10--70\,km~s$^{-1}$.}
\label{Fig:HARP_CO32_spim}
\end{figure*}

\begin{figure}
\centering
\includegraphics[width=8.4cm,clip]{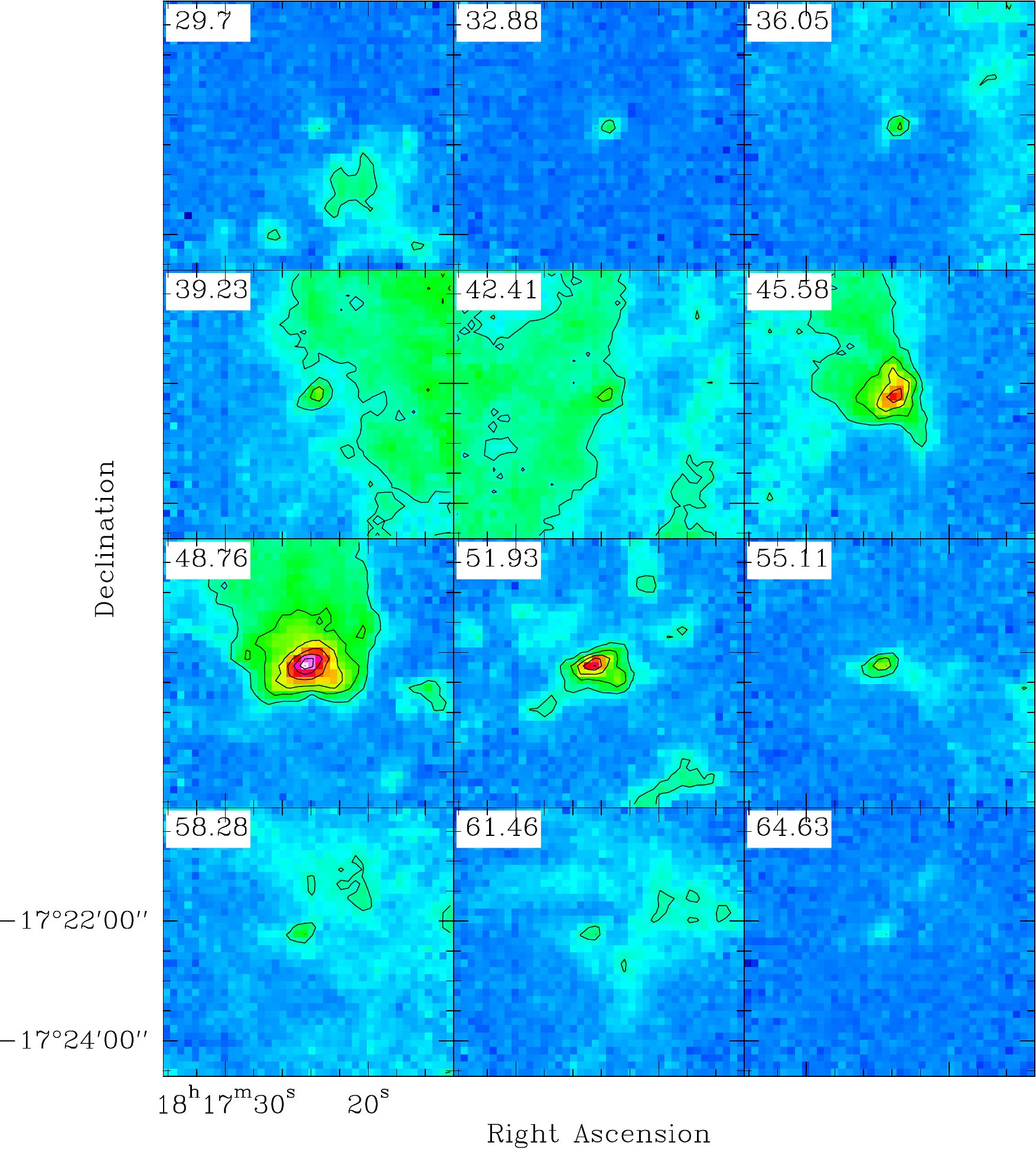}
\caption{Channel maps of the $^{12}$CO (3--2) spectral cube
shown in Fig. \ref{Fig:HARP_CO32_spim}. The maps cover
a field of 5\arcmin$\times$5\arcmin\ and a radial velocity range of 29.7--64.6\,km~s$^{-1}$.
The central velocity of each channel is shown on the uppper left.
The contour levels are 1.5--9\,K in steps of 1.5\,K(T$_A^*$).  }
\label{Fig:HARP_CO32_chmap}
\end{figure}

\begin{figure}
\centering
\includegraphics[width=8.4cm,clip]{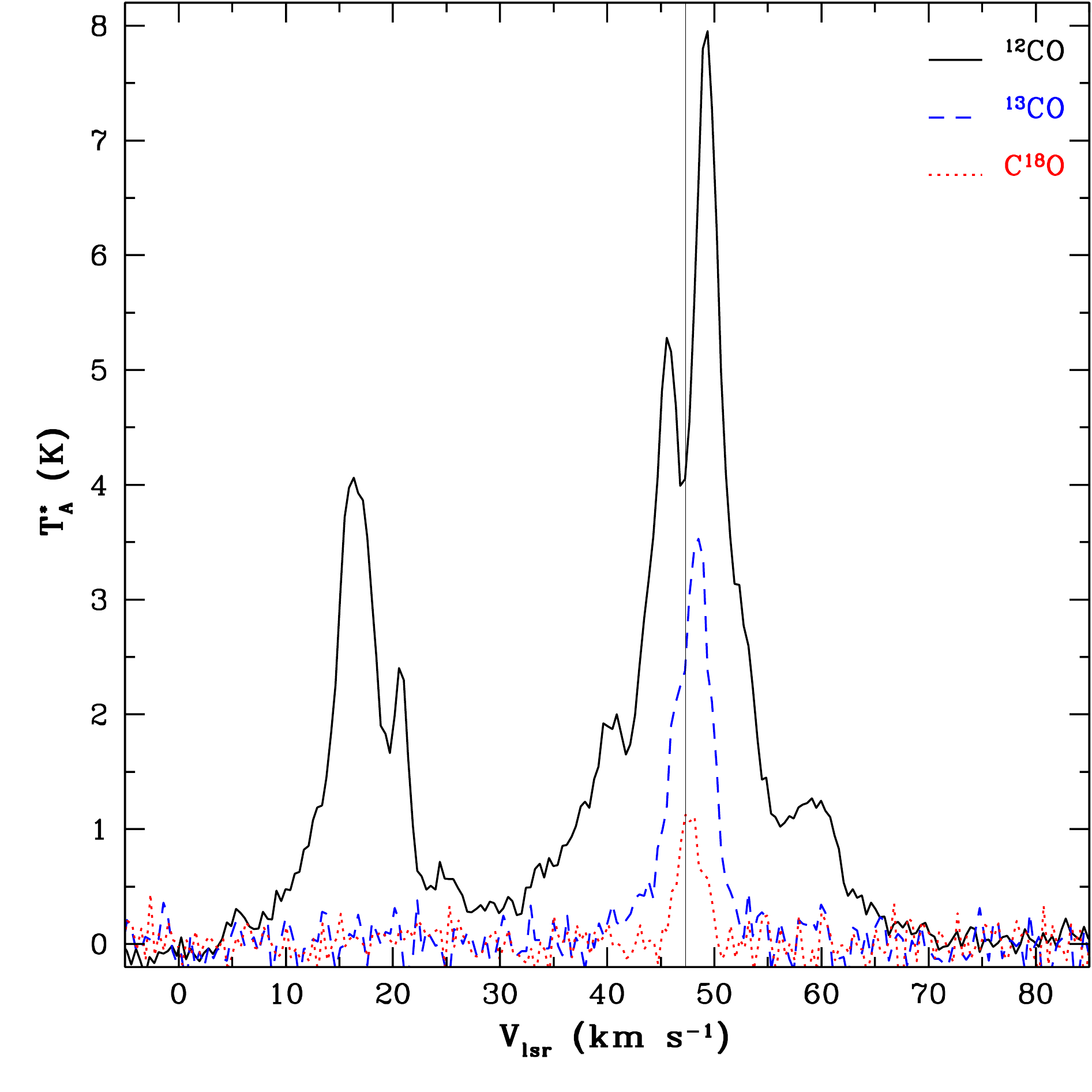}
\caption{The CO(3--2) spectra extracted and averaged in a
43.7\arcsec$\times$43.7\arcsec field.
The solid black line shows the $^{12}$CO and
the dashed blue line shows the $^{13}$CO spectra, both from our
observations on 20140731~UT. The dotted
red line shows the C$^{18}$O spectrum; the archival data
observed on 20080706~UT is presented for C$^{18}$O as it has
lower noise.  The
vertical line shows the systemic velocity of 47.3~km~s$^{-1}$.
The peak T$_{mb}$ observed for the $^{12}$CO line
was 23.09\,K (T$_A^*$=14.78\,K).}
\label{Fig:HARP_CO32_sp}
\end{figure}

Our observations detect a bipolar CO outflow in this 
region for the first time. Fig. \ref{Fig:HARP_CO32_spim} 
shows a section extracted from the $^{12}$CO spectral 
datacube observed on 20140731~UT showing the outflow 
and the emission from the ambient and line-of-sight 
regions.  A 5\arcmin$\times$5\arcmin field in a radial 
velocity range of 10--70\,km~s$^{-1}$ is shown in the 
figure. The red square shows a plane at 47.3\,km~s$^{-1}$, 
the line-of-sight radial velocity of IRAS~18144.
Fig. \ref{Fig:HARP_CO32_chmap} shows a channel map 
of the same area, in a radial velocity range of 29.7--64.6\,km~s$^{-1}$.
From the integrated red-wing, we derive an average 
outflow diameter of 43\arcsec at 3\,$\sigma$ above the 
background level. Therefore, for the purpose of our 
calculations a spectral cube of 6$\times$6 pixels
(43.7\arcsec$\times$43.7\arcsec) dominated by emission 
from the central source and the outflow is used.
Fig. \ref{Fig:HARP_CO32_sp} shows the $^{12}$CO, $^{13}$CO 
and C$^{18}$O(3--2) spectra averaged over this region.  
The $^{12}$CO emission from the core is optically
thick near the line centre and suffers from self absorption.
The contours generated from the blue- and red-shifted 
lobes of the outflow mapped in CO (after integrating in
the velocity ranges given below) are overplotted on 
the continuum-subtracted H$_2$ image in Fig. \ref{Fig:JHH2_cs}.
The CO outflow is in the E-W direction and it traces
the outflow detected in H$_2$, with the blue-shifted lobe of
the CO outflow seen upwind of the MHO~2302 bow-shaped feature,
which probably marks the end of the flow lobe. 

Figs. \ref{Fig:HARP_CO32_spim}, \ref{Fig:HARP_CO32_chmap} and \ref{Fig:HARP_CO32_sp}
reveal the outflow centered at the systemic
velocity, and the emission from the ambient and line-of-sight
clouds. Assuming that the emission close to the line centre is
mostly contributed by the core, to obtain the emission from the
outflow, we integrated the $^{12}$CO(3-2) spectra from
23.3 to 44.0\,km~s$^{-1}$ on the blue wing, and from
50.6 to 71.3\,km~s$^{-1}$ on the red wing of the outflow.
The broad feature with emission peaks centered at 16.1, 20.8 and
25.5\,km~s$^{-1}$ appear to be from
the line of sight clouds, and the features at 40 and
58.5\,km~s$^{-1}$ are due to emission and absorption by
line-of-sight clouds. These features were interpolated
out before integrating the $^{12}$CO(3-2) spectrum to estimate
the emission from the outflow line wings. For 
the velocity range of integration for estimating the emission from the outflow lobes, we set
the velocity of the $^{13}$CO spectrum where the emission
drops to less than 25\% of the intenstity at the peak
as the inner limit.
The outer limits of integration is set where the $^{12}$CO
emission merges with the noise.  The integrated intensities
from the blue- and red-shifted lobes of the outflow
are given in columns 2 and 3 respectively of Table \ref{tab:colden}.

The optical depth in $^{12}$CO ($\tau_{12}$) is often estimated
from a ratio of the antenna temperatures (T$^{*}_{A}$) obtained 
in $^{12}$CO and $^{13}$CO asuming $^{12}$CO/$^{13}$CO = 89
\citep{Garden91}.
Fig. \ref{Fig:HARP_CO32_sp} shows that most
of the emission from $^{13}$CO is from the core. Any 
extended line wing emission in $^{13}$CO is hard to measure due
to noise, which makes the estimate of $\tau_{12}$ from the 
line $^{12}$CO/$^{13}$CO ratio highly uncertain. We, therefore,
assumed optically thin $^{12}$CO emission in line wings and
adopted the limit of $\tau\,\rightarrow$\,0 in the line wings
of the $^{12}$CO emission.
Following \citet{Garden91} and \citet{Varricatt13},
the $^{12}$CO column density is then derived using the equation
\begin{equation}
N_{CO}=\frac{2.39\times10^{14}}{16.6} \frac{e^{\frac{hBJ(J+1)}{kT\textsubscript{ex}}}}{(J+1)} \frac{(T\textsubscript{ex} + 0.92)}{e^{\frac{-16.6}{T\textsubscript{ex}}}}
\int \frac{T^*_A}{\eta_b} d\upsilon~cm^{-1}
\end{equation}
where {\em B} is the rotational constant,
$J$ is the lower level of the transition, and $T$\textsubscript{ex}
is the excitation temperature.

\begin{table}
\caption{$^{12}$CO~(3--2) line intensities and total H$_2$ column
densitites derived for the blue- and red-shifted lobes of the outflow.}
\label{tab:colden}      
\centering                
\begin{tabular}{lllll}        
\hline\hline                                    \\[-2mm]
UT Date &\multicolumn{2}{c}{$\int T^{*}_{A}$dv (K\,km~s$^{-1}$)} &\multicolumn{2}{c}{Column density(cm$^{-2}$)}  \\
(yyyymmdd)  &Blue    &Red       &Blue 	& Red       \\
\hline
20140731 &20.75	&24.90& 1.48$\times 10^{20}$ & 1.78$\times 10^{20}$\\ 
20080419 &20.33 &23.0&  1.45$\times 10^{20}$ & 1.65$\times 10^{20}$\\
\hline
\end{tabular}
\end{table}

Adopting 30\,K for $T$\textsubscript{ex}
\citep{Beuther02b} and $10^4$ for H$_2$/CO, we estimate
1.46$\times$10$^{20}$~cm$^{-2}$ and 1.71$\times$10$^{20}$~cm$^{-2}$
respectively for the average column density of H$_2$ in the blue- and
red-shifted lobes.
Table \ref{tab:colden} gives the column densities derived for the red- and 
blue-shifted lobes of the outflow observed on the two epochs.

The mass in the outflow is estimated using the relation 
$m \times n_{H_2} \times S$, where m is the mean atomic weight,
which is 1.36 $\times$ mass of the hydrogen molecule, and S 
is the surface area of the outflow lobe \citep{Garden91}. As 
the $^{12}$CO column densities were derived from the spectra 
averaged over a 43.7\arcsec$\times$43.7\arcsec field, we 
multiplied the column  densities in the two lobes with the same 
area. Adopting a distance of 4.33\,kpc, we estimate outflow masses 
of 2.67\,M$_\odot$ and 3.13\,M$_\odot$ respectively in the blue- 
and red-shifted wings, and a total mass of 5.8\,M$_\odot$ in the 
outflow. Using a mean velocity outflow $v_{mean}$=13.65\,km~s$^{-1}$
with respect to the v$_{lsr}$ of the core 
for both red- and blue-shifted lobes in the wavelength range of the
outflow lobes considered here,
we estimate an outflow momentum of 79.5\,M$_{\odot}$~km~s$^{-1}$
and energy of 1.08$\times$10$^{46}$\,ergs. The mass we estimate
in the outflow is within the range, but towards the lower end
of the outflow mass typically estimated for massive YSOs 
\citep{Beuther02b, Zhang05}.

Note that our estimate of the mass 
in the outflow should be treated as a lower limit only.  The
calculations are based adopting $\tau \rightarrow$ 0. An 
increase in $\tau$ to 10 will inrease the mass, momentum and 
energy of the outflow by about a factor of 10.

\subsubsection{From the near-IR data}
\label{near-IR}
In the 2.122\,$\mu$m H$_2$ line (Fig. \ref{Fig:JHH2_cs}), 
we obtain a deeper detection of the outflow knot
MHO~2302 previously detected by \citet{Varricatt10}. 44\,GHz 
class~I methanol maser emission is a good tracer of outflows
from massive YSOs. Observations by \citet{Gomez-Ruiz16}
detected 11 maser spots from this region. Nine of those 
are located within 5\arcsec of the H$_2$ knot MHO~2302 and
agree well with the position of the blue shifted lobe of the outflow detected
by us in CO. Note that these maser spots are at velocities close 
to the v$_{lsr}$ of the core; this suggests that these
masers trace post-shock gas in the outflows \citep{Gomez-Ruiz16}.

The spectrum of the outflow lobe MHO~2302 
is given in the lower panel of Fig. \ref{Fig:UIST_sp_csjet}.
Table \ref{tab:H2_lines} gives the intensities 
measured by fitting Gaussians to the lines.
The near-IR spectrum of the outflow shows only the emission lines 
from molecular hydrogen. Except for the 2-1 S(1) line at 
2.24\,$\mu$m, we don't see any emission lines from the vibrational 
states of H$_2$ higher than $v$=1  
suggesting a low level of excitation. Line intensities in
Table \ref{tab:H2_lines} give a low value of
0.15 for $\frac{2-1 S(1)}{1-0 S(1)}$; this low value can
originate from 
either non-dissociative shock-excited gas, fluorescently
excited low-density gas, or UV heated high density (10$^4$--10$^5$~cm$^{-1}$) gas
\citep{Gredel94, Sternberg89}. However, emission from
fluorescent excitation will show H$_2$ from rotational
levels with $\upsilon>$3 \citep{Sternberg89}. While the low line ratio alone will
not let us conclude that the H$_2$ emission is shock-excited, lack of
any high-$\upsilon$ lines strongly suggest a shock-excitation as the emission 
mechanism. In addition, we
don't see the 1.644\,$\mu$m Fe{\sc{II}} line in the outflow which 
suggests a low level of ionization and that the $H_2$ emission 
lines detected are from magnetically-cushioned low-ionization 
C-type shocks rather than dissociative J-type shocks \citep{Oconnel05}. However, note that
the extinction also is high (A$_V\sim$15.6), so we cannot rule out
a non-detection due to extintion.

Under local thermodynamic equilibrium, the column density N($\upsilon$,J) 
is proportional to the statistical weight $g$ and the Boltzman's 
factor $e^\frac{-E(\upsilon,J)}{kT_{ex}}$, where E($\upsilon$,J) is the excitation 
energy, $T_{ex}$ is the excitation temperature and $k$ is the 
Boltzmann's constant. If the rotational and vibrational states are
thermally coupled, we should get a straight line with a slope 
$T_{ex}^{-1}$ when we plot $ln(\frac{N(\upsilon,J)}{g})$ 
against the excitation energy (excitation digram). Following 
\citet{Gredel94}, the column density can be estimated using the 
equation:
\begin{equation}
    I(\upsilon,J) $=$ \frac{hc}{4\pi\lambda}\,A(\upsilon,J)\,N(\upsilon,J)
\end{equation}
where I($\upsilon$,J) is the specific intensity of the line, $h$ is the Plank's constant
and A($\upsilon$,J) is the Einstein's spontaneous emission coefficient.
I($\upsilon$,J) have been extinction corrected following the near-IR extinction
law derived by \citet{Cardelli89}, and adopting R$_V$=A$_V$/E(B-V)=5.0.
Foreground interstellar extinction can also be determined by by iteratively varying A$_V$ and
minimizing the scatter around the straight line fit and thus maximizing
the goodness of the fit.

\begin{figure}
\centering
\includegraphics[width=8.4cm,clip]{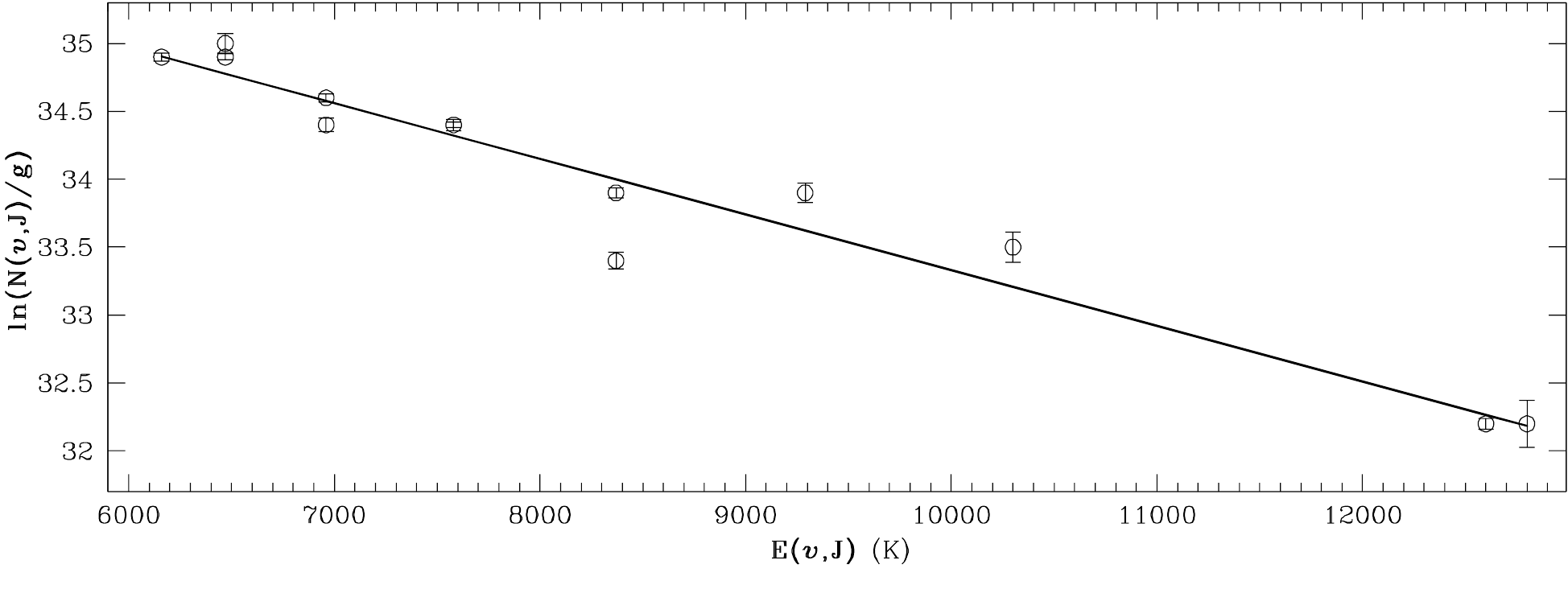}
\caption{The excitation diagram for the $H_2$ lines measured from the
spectrum of MHO\,2302. The open circles are from the measured line intensities
after extinction correction. The straight line shows a  linear fit to the data points.
}
\label{fig:explot}
\end{figure}

Fig. \ref{fig:explot} shows the excitation diagram plotted 
for the lines detected in our spectrum. The straight line fit 
yields a shock excitation temperature of 2440\,K. 
This value is similar to the excitation temperatures derived for
shocked H$_2$ emission for outflows from massive YSOs \citep{Davis04, Caratti15}.
We derive a
foreground extinction $A_V$=15.6 towards the H$_2$ emission feature.

\subsection{The central sources and their natal cloud}
\label{cs_nc}

\begin{figure}
\centering
\includegraphics[width=8.4cm,clip]{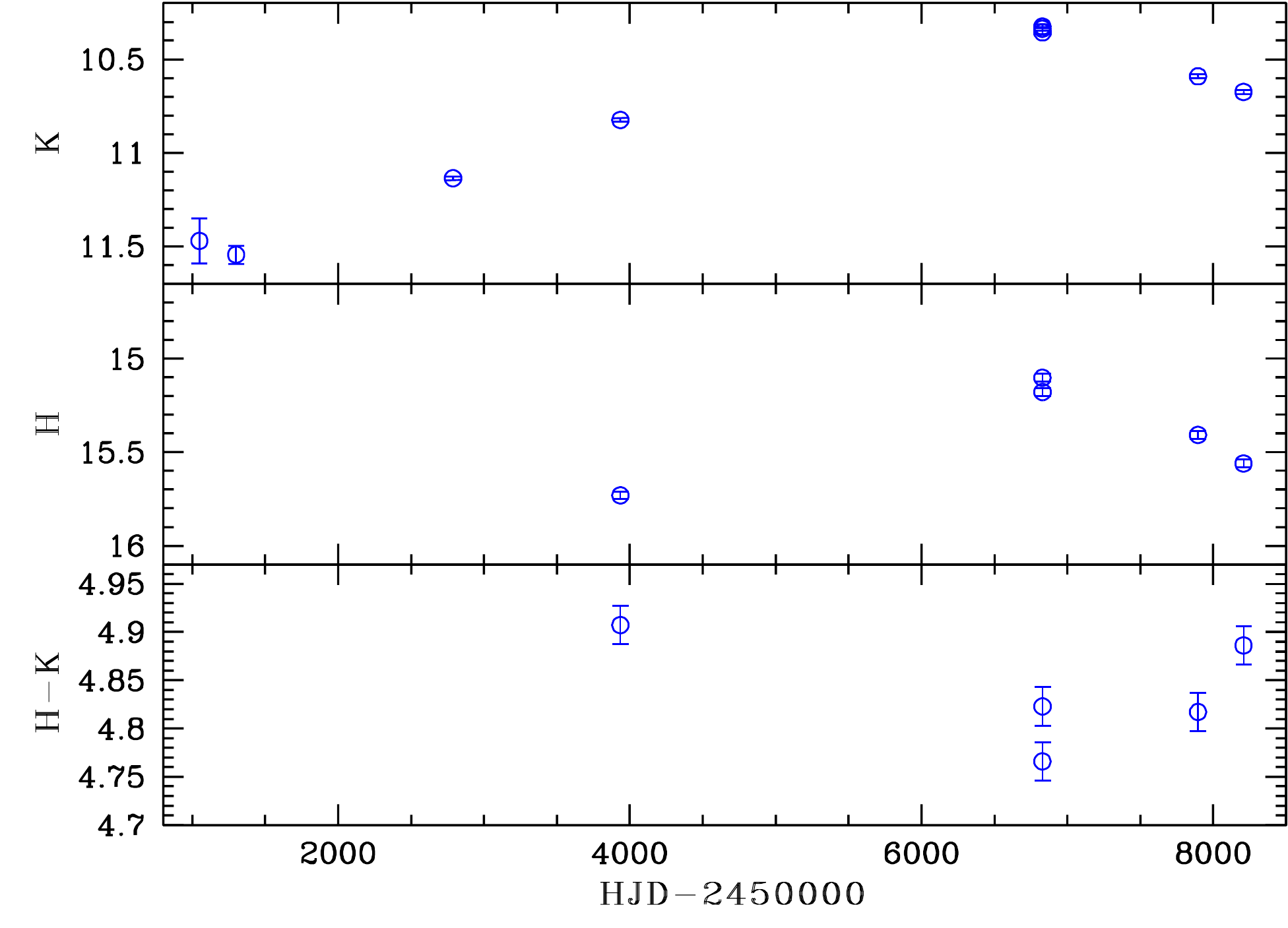}
\caption{$H$ and $K$ magnitudes, and  the
$H-K$ colours of source `A', plotted against the
Heliocentric Julian Date of observations.}
\label{Fig:srcA_K}
\end{figure}

Our high angular resolution infrared observations reveal for 
the first time that the central source of IRAS~18144 is double.
The two sources are are labelled `A' and `B' in 
Figs. \ref{Fig:18144_UIST} and \ref{Fig:18144_Mich}. Source `A' was
the object previously known as the YSO driving the outflow \citep{Varricatt10}.
It is deeply embedded and is
detected well in $K$ band only, with a marginal
detection in $H$. Table \ref{tab:SrcA_K} shows a compilation of
the $K$-band photometry available for this source. In $H$ band,
we detect a faint point source associated with a cometary
nebula. We estimated the photometry in $H$ band using a 
1.2\arcsec-diameter aperture to avoid as much of the contribution
from the nebula as possible, and applying aperture correction.
The $H$-band magnitudes are also presented in
Table \ref{tab:SrcA_K}. Fig. \ref{Fig:srcA_K} shows the 
variations in the $H$ and $K$ magnitudes and the 
$H$-$K$ colours.
We see that source `A' brightened by $\sim$1.2 mag in $K$ 
during a period of 15 years from 1999 to 2004. Our most recent observations
in 2017 and 2018 show that the `A' is fading now. 
A search by \citet{Pena14} in the 
UKIDSS GPS (Galactic Plane Survey) data of a limited 
area of the surveyed region with two epochs of $K$-band 
photometry available suggested that stellar variability 
with $\Delta\,K$ > 1 mag is very rare.
They proposed that most of these high amplitude variables are YSOs.
Eruptive YSOs have been classified into FU Ori variables, which
exhibit a fast brightening associated with sudden increase in accreation 
rate followed by a slow decline over a period of 
over 10 years, or staying in the stage of elevated accretion 
for decades, and Exor variables, which have recurrent 
short-lived ($<$1 year) outbursts separated by a few years of
quiescence \citep{Pena14,Audard14}. Both these phenomena
are seen in low luninosity objects.
With the brightening
happening during $\sim$15 years, source `A' does not fall 
into the category of currently classified eruptive pre-main sequenced variables. 
Fig. \ref{Fig:srcA_K} shows that the reddening is lowest when
the source is brightest in $H$ and $K$ suggesting that 
variable amount of obscuration
by dust is a major contributor to the variability.

Table \ref{tab:Mich_imaging_log} shows the flux densities of sources
`A' and `B' measured in different Michelle filters. `A' shows
variability well above the 1$\sigma$ error limits in all Michelle filters.
The variations are suggestive of a fading from 2004 to 2016 and a
re-brightening. However, these variations are less
than 3$\sigma$. `B' does not exibit any variability above
the 1$\sigma$ error limits. 

\begin{table}
\centering
\begin{minipage}{140mm}
\caption{$H$ and $K$-band magnitudes of source `A'}
\label{tab:SrcA_K}
\begin{tabular}{@{}llllll@{}}
\hline
Data     &UT   &\multicolumn{2}{c}{H}	&\multicolumn{2}{c}{K}	\\ 
source	 &     &mag       &error	&mag	&error\\
\hline
DENIS     &19980823.06580    &&&11.47$^{\mathrm{c}}$ &0.120\\    
2MASS     &19990502.36941    &&&11.55$^{\mathrm{c}}$ &0.049\\
UFTI	  &20030528.57279    &&&11.136$^{\mathrm{d}}$ &0.005\\
WFCAM$^{\mathrm{a}}$  &20060718.46093   &15.46  &0.02 &&\\
WFCAM$^{\mathrm{a}}$  &20060718.46600   &&&10.824 &0.001\\
WFCAM$^{\mathrm{b}}$  &20140620.44216    &14.93  &0.02 &&\\
WFCAM$^{\mathrm{b}}$  &20140620.44627    &&&10.337 &0.010\\
WFCAM$^{\mathrm{b}}$  &20140620.44822    &&&10.324 &0.010\\
WFCAM$^{\mathrm{b}}$  &20140621.45545    &14.94   &0.02 &&\\
WFCAM$^{\mathrm{b}}$  &20140621.46008    &&&10.356 &0.010\\
WFCAM$^{\mathrm{b}}$  &20170522.56375    &15.18    &0.02 &&\\
WFCAM$^{\mathrm{b}}$  &20170522.56840    &&&10.591 &0.010\\
WFCAM$^{\mathrm{b}}$  &20180330.62079   &15.29    &0.02 &&\\
WFCAM$^{\mathrm{b}}$  &20180330.62546   &&&10.675 &0.010\\
\hline
\end{tabular}
\end{minipage}
\begin{list}{}{}
\item[$^{\mathrm{a}}$]From the UKIDSS survey; $^{\mathrm{b}}$This work; $^{\mathrm{c}}$After conversion to UKIRT-WFCAM system using the $J$-band detection limit of our WFCAM observations; $^{\mathrm{d}}$Derived from the $K$-band image of \citet{Varricatt10}
\end{list}
\end{table}

\begin{figure}
\centering
\includegraphics[width=8.4cm,clip]{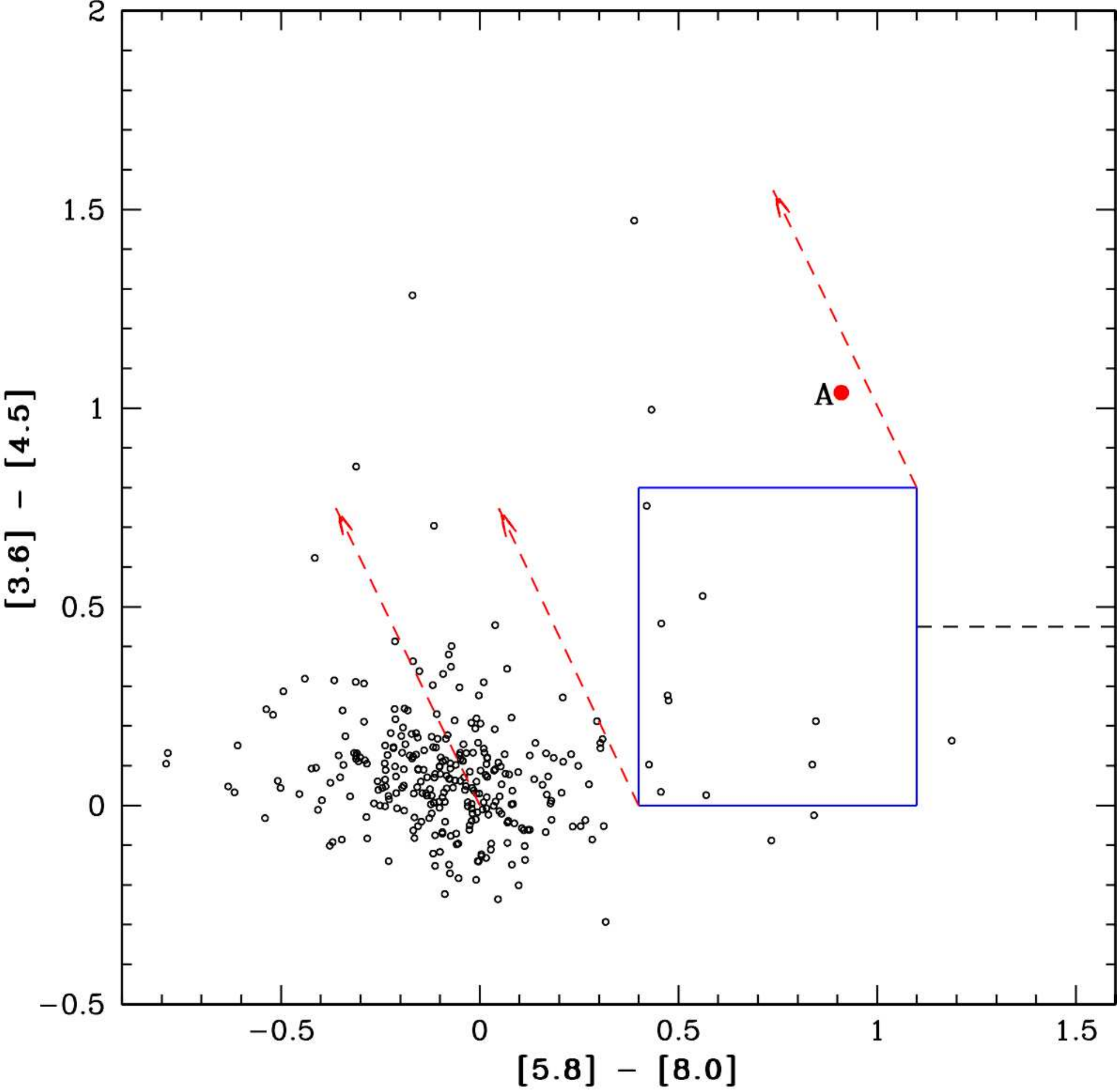}
\caption{Colour-colour diagram of the objects detected in
all IRAC bands in a 10\arcmin$\times$10\arcmin\ field centred
on IRAS~18144 (open circles). The filled red circle shows
source `A'. The blue rectangle shows the approximate location
of Class II sources, and the dashed horizontal line shows
the boundary between Class I/II objects (below) and
Class I objects (above). The dashed red arrows show
reddening vectors for A$_V$ = 45.
}
\label{Fig_IRAC_col}
\end{figure}

{\it Spitzer} IRAC colours can be used to classify YSOs
using the scheme defined by \cite{Allen04} and \cite{Megeath04}.
IRAC
([5.8]-[8.0], [3.6]-[4.5]) colours of objects detected in a
10\arcmin$\times$10\arcmin\ field around IRAS~18144 are
plotted in Fig. \ref{Fig_IRAC_col}.  A majority
of the detections are around ([5.8]-[8.0], [3.6]-[4.5]) = 0.
These are mostly foreground and background stars and diskless
pre-main-sequence (Class III) objects. The blue rectangle
(0 < ([3.6]-[4.5]) < 0.8 and 0.4 < ([5.8]-[8.0] < 1.1) shows
the approximate location of Class II objects, and the dashed
horizontal line shows the boundary between Class I/II (below) 
and Class I objects (above). The sources with ([3.6]-[4.5]) > 0.8
and ([5.8]-[8.0]) > 1.1 are likely to be Class I objects, which 
are protostars with infalling envelopes.
The dashed red arrows 
show reddening vectors for A$_V$ = 45 calculated from the 
extinction law given in Mathis (1990). 
Sources `A' and `B' are not resolved well in
all {\it Spitzer} bands, and `B' is not detected in bands 1 and 2. 
As the 5.8 and 8.0\,$\mu$m magnitudes of `A' in the 
{\it Spitzer} catalogue may be influenced by the close proximity
of `B', we performed aperture phtometry of `A' from the IRAC
images in these
two bands, using the mean zero points estimated from a few 
isolated point sources in the field 
and adopting a 3\arcsec-diamter aperture. We derive  
magnitudes of 4.95$\pm$0.06 and 4.04$\pm$0.05 respectively in the 
5.8 and 8\,$\mu$m bands, which are used to derive its
[5.8]-[8.0] colour. 
The location of source `A' in Fig. \ref{Fig_IRAC_col} suggests that it is 
very young, and is likely to be a reddened
Class II object. 
With `B' being fainter and  not resolved well in the {\it Spitzer} images, it is 
difficult to place it in this diagram.  Using its
flux density measured in the UIST $M'$ and
Michelle 7.9\,$\mu$m bands, approximating the SED 
with a linear relation, and using the {\it Spitzer}-IRAC
zero points, we estimate [5.8]-[8.0]$\sim$1.8. This
will place `B' in the region of Class I or
Class I/II sources in Fig. \ref{Fig_IRAC_col}.

The $HK$ spectrum of source `A' (the upper panel of 
Fig. \ref{Fig:UIST_sp_csjet}) shows a rising SED, 
free of photospheric absorption lines suggestive of a 
YSOs at high extinction. The prominent spectral feature 
is the Br$\gamma$ emission; we measure an integrated flux
of 1.36$\times$10$^{-17}$\,W\,m$^{-2}$ and an equivalent
width of -6.72\,\AA\ for this line. Assuming that the 
Br$\gamma$ line emission is due to accretion, we can use 
the line strength to obtain an approximate estimate of 
the accretion rate. The observed line flux was extinction 
corrected assuming A$_V$=15.6 towards MHO~2302 derived
in \S\ref{near-IR} as a lower limit for the extinction
towards source `A', and employing the
near-IR extinction law derived by \citet{Cardelli89}.
A value of 5 was adopted for R$_V$.
From the extinction corrected flux,
we derive a luminosity 
of 5.51$\times$10$^{-2}$~L$_{\odot}$ in Br$\gamma$, for a distance of 
4.33\,kpc. 
This gives accretion luminosity of 10.2\,L$_{\odot}$, 
using the Eq. 2 of \citet{Fairlamb17}.
Employing Eq. 11 of \citet{Hartigan03}, and using 
M$_{\star}$=18\,M$_{\odot}$ and R$_{\star}$=6.3\,R$_{\odot}$
from \citet{Grave09} and the accretion luminosity, we estimate 
an accretion rate of 2.2$\times$10$^{-5}$\,M$_{\odot}$\,y$^{-1}$
for source `A'.
Note that this value can have large uncertainty as the
A$_V$ could be different towards source `A', and
the values of M$_{\star}$ and R$_{\star}$ estimated by 
\citet{Grave09} could be different from the true 
values for source `A' as the flux densities used in their
SED analysis were the combined values for source `A' and `B'.

`B' is detected as a deeply embedded source with strong 
IR colours, and is visible in our images only in $M'$ band
and at longer wavelengths. Its $M'$-band detection is only 
marginal. It is detected well at longer wavelengths with 
Michelle, and the Michelle photometry shows a deeper 
10\,$\mu$m Silicate absorption band than for `A' 
(Table \ref{tab:Mich_imaging_log}), suggesting that `B' 
is in a more deeply embedded phase than `A'. 
Source `B' does not exhibit any mid-IR variability above 1$\sigma$.
Both `A' and `B' are likely to be YSOs, with
`B' in a much younger phase than `A'. 
`B' is probably a Class I or Class I/II object in the envelope phase, 
where `A' seems to be more advanced Class II object, 
which has cleared of most of the envelope.  
With the appearance of `B' as the younger of 
the two YSOs detected here, its location
closer to the centroid of the outflow lobes, and the
location of the SCUBA2 and Herschel sources closer to
`B' than to `A', we conclude that `B' is the YSO 
responsible for the outflow seen in CO and H$_2$, 
and that it hosts an accretion disk.

The CO(3--2) data can be used to estimate a lower limit 
to the mass of the cloud. Our CO(3--2) datacube observed
on 20140731 UT was integrated in the 32--55~km~s$^{-1}$ velocity 
range, the integrated line emission over the whole cloud
was extracted and was corrected to the main beam brightness 
temperature using a main beam efficiency of 0.64 for HARP.
We assumed a CO-H$_2$ conversion
factor (X-factor) of 2.0$\times$10$^{20}$\,cm$^{-2}$\,/\,K~kms$^{-1}$ \citep{Bolatto13}
and a ratio of CO(1--0) to CO(3--2) main beam brightness temperatures
of 1.0 (the X-factor is defined for CO J=1--0).
We derive a total mass of 3994~M$_\odot$ for the cloud associated with IRAS~18144.
The mass estimate should be treated only as a lower limit to 
the mass of the cloud as the CO can be underestimated due
to optical thickness and feeze out on to dust grains.

The detection of HCO$^+$(4-3) (\S\ref{HCO+}) alone does not allow for
an accurate determination of temperature and density in
the cloud. However the energy of the $J$=4 level is
43\,K above the ground state and therefore it is likely
that the kinetic temperature is higher than $\sim$20\,K.
\citet{Molinari96} derive a kinetic temperature of
23.61\,K from NH$_3$ observations. \citet{Evans99} lists
the effective density (the critical density, corrected
for trapping) needed for the $J$=4-3 transition. It is
5.0$\times$10$^5$\,cm$^{-3}$ for a T$_{kin}$ of 10\,K
and 10$^4$\,cm$^{-3}$ for 100\,K. The densities are therefore
expected to be around 10$^5$\,cm$^{-3}$.

The HCO$^+$ emission is slightly elongated EW, with a position
angle of 77$^\circ$ and an FWHM of 25.5\arcsec$\times$19.5\arcsec.
This gives a mean radius of 0.24~pc for a distance of 4.33~kpc.
After averaging over a 43.7\arcsec$\times$47.3\arcsec area over which
the emission from the core is detected in HCO$^+$, we get an
FWHM of 4.7\,km~s$^{-1}$ for the line.
We can calculate the virial
mass of the core following the method adopted by \citet{Maclaren88} 
as M=k\,R\,$\Delta$V$^2$, where M is in
solar mass, R is the radius in parsec and $\Delta$V is the
FWHM of the line in km~s$^{-1}$. The value of k is 190 and 126 for a
density distribution of r$^{-1}$ and r$^{-2}$ respectively within the core.
This gives us a virial mass of 1007~M$_\odot$ for
an r$^{-1}$ density distribution, and 668~M$_\odot$ for an
r$^{-2}$ density distribution.

We derive the mass of the cloud from the 850\,$\mu$m
SCUBA-2 map following \citet{Deharveng09}, using a
gas-to-dust ratio of 100. Assuming a dust temperature
of 23.61\,K, the kinetic temperature derived by 
\citet{Molinari96} from NH$_3$, and a total flux
density of 11.8\,Jy at 850\,$\mu$m (Fig. {\ref {tab:SCUBA2_flx}}), we obtain a mass of
883\,M$_\odot$. This mass is consistent with the virial
mass derived from the HCO$^+$(4--3) map, but smaller than the mass
derived from $^{12}$CO(3--2) which is more extended.
The core and cloud mass we derive for IRAS~18144
are comparable to the values reported towards other
massive YSOs in previous studies \citep{Beuther11, Mookerjea07}.

\subsubsection{The spectral energy distribution}
\label{fitting}

The SED of the central source was modelled using
the SED fitting tool of \citet{Robitaille07}.
This tool uses a grid of 2D radiative transfer models presented
in \citet{Robitaille06}, and developed by \citet{Whitney03a} and
\citet{Whitney03b}.  The grid is composed of 20,000 YSO models,
and covers objects in the 0.1--50\,M$_\odot$ range in the
evolutionary stages from early envelope infall stage to the
late disk-only stage. Each model is available at  10 different
viewing angles.  

\begin{figure}
\centering
\includegraphics[width=8.5cm,clip]{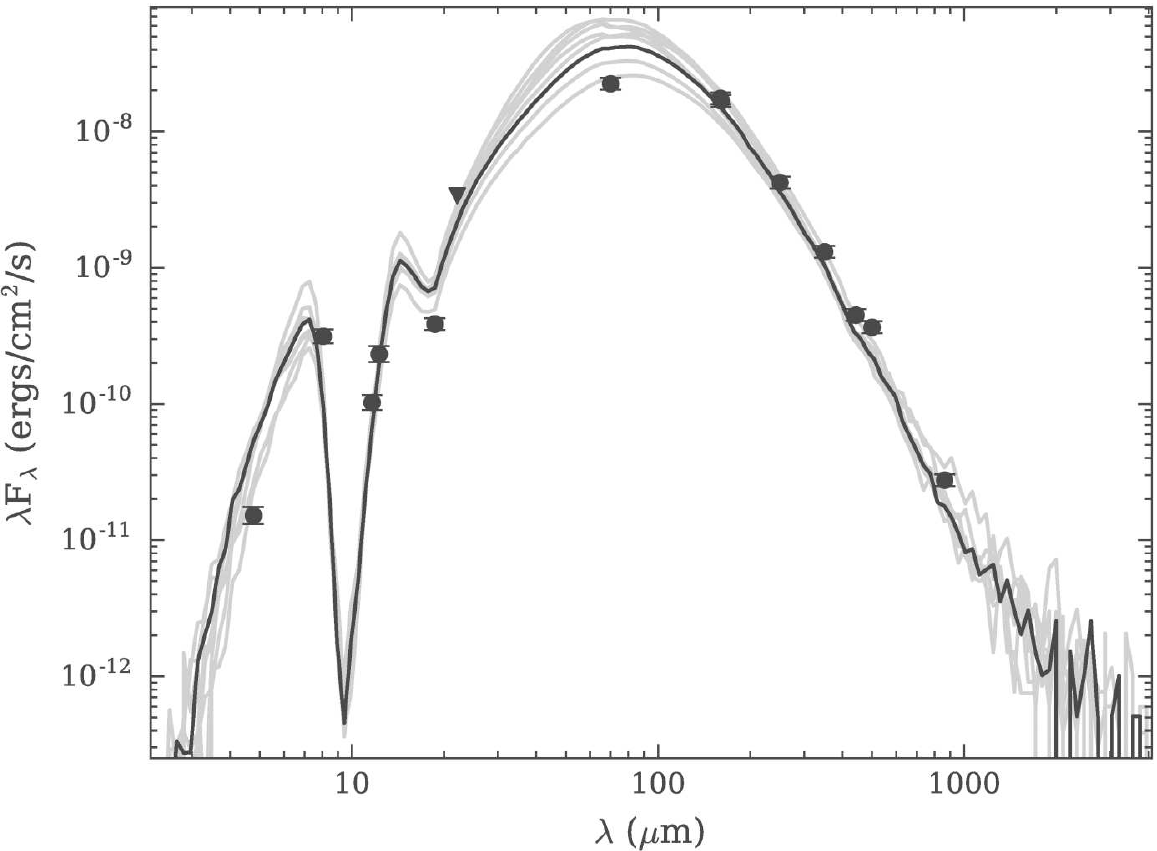}
\caption{The black line shows the best fit model to the 
data points. The grey line shows the only other model 
with $\chi^2-\chi^2_{best fit}$ per data point $<$ 3.
}
\label{Fig:IRAS18144_SED}
\end{figure}

The fit was performed assuming `B' as the sole 
contributor to the flux densities used.
We used the flux density measured in the UIST $M'$ band 
and in the the four Michelle bands. The WISE band W4 data
was used, but only as an upper limit as `A' 
will still be a significant contributor at 22\,$\mu$m.
AKARI-FIS 65 and 160\,$\mu$m fluxes, Herschel measurements 
from Table\,\ref{tab:Herschel_flx} and our SCUBA2 measurements 
from Table \ref{tab:SCUBA2_flx} were also used for fitting 
the SED. A mininimum uncertainty of 10\% of the observed 
flux densities was used for observations with smaller 
uncertainties.
Fig. \ref{Fig:IRAS18144_SED} shows the fit to the 
SED.  The solid black line shows the best fit model. The 
grey lines show the models with
$\chi^2-\chi^2_{best fit}$ per data point $<$ 3. 
Our analysis shows a 16.2\,M$_{\odot}$
star with a disk accretion rate of 9$\times$10$^{-9}$\,M$_\odot$/year,
envelope accretion rate of 4.25$\times$10$^{-3}\,$M$_\odot$/year,
total luminosity of 3.07$\times$10$^{4}\,$L$_\odot$ and
a foreground extinction A$_V$ of 28.7 mag. Table \ref{tab:IRAS18144_SED} shows the parameters of the
best fit model.

SED analysis by \citet{Grave09},
derived M=18\,M$_\odot$, age=1.6$\times$10$^5$\,years, T$_{*}\sim$20,000\,K,
M$_{disk}\sim$0.08\,M$_{\odot}$, and disk and envelope accretion rates
of $\sim$4$\times$10$^{-6}$\,M$_\odot$ and $\sim$6.4$\times$10$^{-5}$\,M$_\odot$
respectively. 
Similarly the SED analysis of \citet{Tanti12} also yielded
a 15.44\,M$_\odot$ YSO with age=2.1$\times$10$^4$\,years,
T$_{*}\sim$14,321\,K, and
a luminosity of 2.22$\times$10$^4$\,L$_{\odot}$. Sources `A' and
`B' were not resolved in these studies. Both these studies and 
our analysis used the SED fitting tool of \citet{Robitaille07},
with the same underlying models and pre-main-sequence evolutionary tracks.

Whereas the previous studies may have been affected by poor 
spatial resoltion at all wavelengths,  poor resolution at 
far-IR and sub-mm wavelengths are likely to be affecting 
our results. Even though the near and mid-IR colours 
show that `B' as the younger source with a 
steeper SED when compared to `A', the disk accretion rate 
for `B' derived from the SED analysis is lower than the 
the accretion rate for `A' we derive from the 
Br$\gamma$ luminosity. If `B' is comparable to `A' in mass
and is the younger one driving the outflow seen in CO, we 
expect a larger accretion rate for it. 

In general, all disk parameters estimated by the 
SED fit for `B' have large uncertainty. Note that at 
near and mid-IR wavelengths, we resolve `A' and `B' 
and the flux densities of `B' are used in the SED modelling.
At longer wavelengths, the spatial resolution is not
sufficient to resolve the two sources. The SED modelling
is based on the assumption that a single source is
responsible for the observed flux densities.
Treating the emission at these
wavelenghts as entirely from `B' is likely to be
significantly affecting the estimation of the disk
parameters. If `A' is
also a YSO with strong accretion, its contribution
in the far-IR and sub-mm wavelengths will not
be negligible. Alternatively, `B' itself may
be composed of more than one source.
We therefore need high angular resolution
observations at longer wavelengths to gain better 
understanding of this system.  It should also
be noted that the disk accretion rate for embedded sources 
may not be well-determined due to the fact that the UV accretion 
luminosity will be reprocessed by absorption and re-emission 
in the envelope, which has the effect of increasing the total 
luminosity of the source while not changing the shape of the SED.

\begin{table}
\caption{Results from SED fitting}             
\label{tab:IRAS18144_SED}      
\centering                      
\begin{tabular}{ll}        
\hline\hline                                    				\\[-2mm]
Parameter                         	&Best-fit values$^{\mathrm{a}}$		\\
\hline                                          				\\[-2mm]
Stellar mass (M$_{\odot}$)              &16.2 (15.6--21.3)                           \\
Stellar radius (R$_{\odot})$            &6.4 (6.4--70.3)                             \\
Stellar temperature (K)                 &3.03 (0.8--3.63)$\times$10$^{4}$           \\
Stellar age (yr)                        &4.09 (0.83--6.60)$\times$10$^{4}$          \\
Disk mass (M$_{\odot}$)                 &4.72 (0.0--133)$\times$10$^{-3}$           \\
Disk accretion rate (M$_{\odot}$yr$^{-1}$) &9 (0.0--167600)$\times$10$^{-9}$      \\
Disk/Envelope inner radius (AU)         &46.7 (0.0--50.4)                            \\
Disk outer radius (AU)                  &5312 (0.0--5312)                            \\
Envelope accretion rate (M$_{\odot}$yr$^{-1}$) &4.25 (2.25--5.59)$\times$10$^{-3}$   \\
Envelope outer radius (AU)              &1.00 $\times$10$^{5}$                         \\
Angle of inclination of the disk axis ($^{\circ}$)  &31.8 (31.8--56.6)               \\
Total Luminosity (L$_{\odot}$)          &3.07 (1.81--6.24)$\times$10$^{4}$           \\
A$_V$ (foreground)                      &28.7 (4.0--41.0)                            \\

\hline
\end{tabular}
\begin{list}{}{}
\item[$^{\mathrm{a}}$]There were seven other models with 
($\chi^2-\chi^2_{best fit}$) per data point $<$ 3.
The values
given in parenthesis are the ranges in the parameter values
for all these models.
A distance of 4.33\,kpc is adopted for IRAS~18144.
\end{list}
\end{table}

\section{Conclusions}

\begin{enumerate}

\item Our sub-mm observations of IRAS~18144 reveal 
massive star formation taking place in a dense core 
located in an isolated molecular cloud.

\item We discover an E--W outflow in the CO(3--2) line. 
The CO outflow is aligned with the jet imaged by us 
in the 2.122\,$\mu$m H$_2$ line from this region showing 
that the outflow is jet-driven. We estimate a lower limit 
of 5.8\,M$_\odot$ for the mass in the outflow.

\item Analysis of the near-IR spectrum of the
jet discovered in H$_2$ gives and excitation temperature
of 2440\,K and a foreground extinction of A$_V$=15.6 mag
for the H$_2$ line emission lobe.

\item Through near- and mid-IR imaging, we discover that 
IRAS~18144 hosts at least two massive YSOs (labelled
`A' and `B') in a dense core mapped at 450 and 850\,$\mu$m. 
The newly discovered
embedded source `B' is younger and is likely to be the one 
driving the outflow seen in CO and H$_2$. Both sources 
appear to be accreting, and are likely to be in class II 
and class I stages respectively.

\item The $HK$ spectrum of source `A' shows a steeply
rising SED, with Br$_\gamma$ emission, which is typical
of a YSO with accretion.  Assuming that the Br$_\gamma$
emission is from the accretion disk, we calculate
an accretion rate of 2.2$\times$10$^{-5}$\,M$_\odot$~year$^{-1}$.

\item Source `B' is resolved well from `A' in our
mid-IR observations. The spatial resolution of the far-IR 
and sub-mm data are not sufficient to resolve the two sources.  
Assuming that `B' is the sole contributor to the emission 
observed at far-IR and sub-mm wavelengths, we modelled
the SED, which revealed a 16.2\,M$_\odot$ YSO with a
luminosity of 3.07$\times$10$^4$\,L$_\odot$. The disk parameters
exhibited from the SED analysis have large uncertainties.
It is likely that our assumption that `B' is the sole contributor
to the observed flux densities at longer wavelengths 
is not justified, and that the contribution from source `A',
at those wavelengths cannot be neglected. Even though the 
two sources appear close at 2.6\arcsec, they have a 
projected separation of 11258\,AU at a distance of 4.33\,kpc. 
We also cannot ignore the possiblity of 
`B' being composed of more than one source, 
which are not resolved in our observations.
High angular resolution observations at far-IR and 
sub-mm wavelengths are required to reliably estimate
the parameters of the YSOs, and to learn if there are
more YSOs associated with this cloud.

\item An examination of the near-IR magnitudes available
for source `A', spanning a period of $\sim$20 years,
show that source brightened by over 1.1 magnitudes and is
now fading. The variation of $H-K$ suggests that the
variability is due to varying amount of obscuration by
circumstellar dust.

\item We conclude that IRAS~18144 hosts at least two 
massive YSOs, located inside a dense core situated in
an isolated cloud. We estimate a virial mass of 1007\,M$_{\odot}$ 
for the core from archival HCO$^+$(4--3) data. This is 
comparable to a mass of 883\,M$_{\odot}$ derived from our
SCUBA-2 850\,$\mu$m observations. The SCUBA-2 continuum 
maps and HCO$^+$(4--3) maps show the densest region of 
the cloud (core) where the two YSOs are located. 
From our $^{12}$CO(3-2) data, we estimate a lower limit
of 3994\,M$_\odot$ for the mass of the cloud hosting the core.

\end{enumerate}

\section*{Acknowledgements}

UKIRT is owned by the University of Hawaii (UH) and operated by 
the UH Institute for Astronomy; the operations are enabled through the 
cooperation of the East Asian Observatory (EAO). When some of the 
data reported here were acquired, UKIRT was supported by NASA and 
operated under an agreement among the UH, the University of Arizona, 
and Lockheed Martin Advanced Technology Center.
JCMT has historically been operated by the Joint Astronomy Centre 
on behalf of the Science and Technology Facilities Council of the 
United Kingdom, the National Research Council of Canada and the 
Netherlands Organisation for Scientific Research.
Additional funds for the construction of SCUBA-2 were provided 
by the Canada Foundation for Innovation.
We use of data obtained with
AKARI, a JAXA project with the participation of ESA.
The archival data from Spitzer, WISE and Herschol are downloaded from NASA/IPAC
Infrared Science Archive, which is operated by the Jet Propulsion Laboratory,
Caltec, under contract with NASA. 
We thank UKIRT and JCMT staff members for helping us with the
observations, and the Cambridge Astronomical
Survey Unit (CASU) for processing the WFCAM data. We thank
the anonymous referee for the comments and suggestions, which have
improved the paper.








\appendix

\bsp	
\label{lastpage}
\end{document}